\documentclass[twocolumn]{aastex63}
\usepackage{hyperref}
\usepackage[colorinlistoftodos]{todonotes}

\newcommand{\msun}{M$_{\odot}$}

\shorttitle{Analysis of SNR W49B}

\begin{document}

\title{Analysis of XMM-Newton Observations of Supernova Remnant W49B and Clues to the Progenitor}

\correspondingauthor{Vikram V. Dwarkadas}
\email{vikram@astro.uchicago.edu}

\author[0000-0002-9337-0902]{Jared Siegel}
\affiliation{Department of Astronomy and Astrophysics, University of Chicago \\
5640 S Ellis Ave \\
Chicago, IL 60637, USA}

\author[0000-0002-4661-7001]{Vikram V. Dwarkadas}
\affiliation{Department of Astronomy and Astrophysics, University of Chicago \\
5640 S Ellis Ave \\
Chicago, IL 60637, USA}

\author[0000-0003-0570-9951]{Kari A. Frank}
\affil{Center for Interdisciplinary Exploration and Research in Astrophysics, Northwestern University \\
1800 Sherman Ave\\
Evanston, IL 60201, USA}

\author[0000-0003-0729-1632]{David N. Burrows}
\affiliation{Department of Astronomy and Astrophysics, Pennsylvania State University \\
525 Davey Laboratory \\
University Park, PA 16802, USA}

\begin{abstract}

W49B is a supernova remnant (SNR) discovered over 60 years ago in early radio surveys. It has since been observed over the entire wavelength range, with the X-ray morphology resembling a  centrally-filled SNR. The nature of its progenitor star is still debated. Applying Smoothed Particle Inference techniques to analyze the X-Ray emission from W49B, we characterize the morphology and abundance distribution over the entire remnant. We also infer the density structure and derive the mass of individual elements present in the plasma. The morphology is consistent with an interaction between the remnant and a dense medium along the eastern edge, and some obstruction towards the west. We find a total mass of 130 $(\pm 16)$ {\msun} and an estimated ejecta mass of 1.2 $(\pm 0.2)$ \msun. Comparison of the inferred abundance values and individual element masses with a wide selection of SN models suggests that deflagration-to-detonation (DDT) Type Ia models are the most compatible, with Fe abundance being the major discriminating factor. The general agreement between our abundance measurements and those from previous studies suggests that disagreement between various authors is more likely due to the choice of models used for comparison, rather than the abundance values themselves. While our abundance results lean toward a Type Ia origin, ambiguities in the interpretation of various morphological and spectral characteristics of W49B do not allow us to provide a definitive classification.

\end{abstract}

\keywords{X-rays: individual (W49B) --- supernovae: individual (W49B) --- nuclear reactions, nucleosynthesis, abundances --- ISM: supernova remnants --- X-rays: ISM --- circumstellar matter}

\section{Introduction}

The core-collapse and explosion of a massive star ($\ga 10 M_{\odot}$), or the thermonuclear deflagration and detonation of a white dwarf, are the leading pathways to a  supernova (SN) explosion. The resulting shock wave continues to expand out into the surrounding medium for thousands of years, sweeping up the surrounding medium to form large structures made of gas and dust, that are observable over the entire wavelength range. These explosions produce most of the elements in the universe besides Hydrogen and Helium, thus making them of great scientific interest. Their large size makes them easily observable, leading to many being detected in the early days of radio astronomy.

The W49 complex was first detected in 1958 by \citet{westerhout58} in a survey of the Galaxy at 22cm. It was soon found to be made of two distinct regions \citep{mst67}, one of which (W49A) had a thermal component, and the other (W49B) a non-thermal component. \citet{mst67} identified the latter as an ``old supernova remnant" in the Galaxy. Further and more resolved radio observations \citep{wilson75, dw74} confirmed the SNR designation. X-rays were first detected by \citet{pye1984}, although the spectral data did not allow the nature of the X-ray emission to be inferred, with both thermal and synchrotron models proving adequate fits. A much higher quality spectrum with {\it ASCA} \citep{fujimotoetal95} showed that the observed spectrum was consistent with thermal emission, and required multiple plasma components. Subsequent detailed analysis of {\it ASCA} data led \citet{hwang2000} to suggest that W49B was a Type Ia remnant, based on the relative element abundances, but they were unable to make a conclusive determination. Using {\it Chandra} and {\it XMM-Newton} data, \citet{lopezetal09} investigated the morphological asymmetries and suggested that the global abundances were consistent with a bipolar explosion from a 25 \msun\ star. In the same year, the discovery of a radiative recombination continuum bump in the {\it Suzaku} spectra \citep{ozawaetal09} showed that the plasma was in a highly overionized state, making W49B one of the first overionized remnants to be identified. Spatially, the overionized component was found to occupy the central region and the `western jet' \citep{miceli2010}. More recently, the abundance ratios and metal distributions in W49B were found to be more consistent with a Type Ia origin for the SNR \citep{zv18}.

\subsection{Multi-wavelength Observations of W49B}
\label{sec:mw} 
W49B lies at an estimated distance of 8-10 kpc \citep{radhak1972,moffett1994,Z14}, with an approximate age of 1000-6000 years \citep{pye1984, smith1985, hwang2000,zv18}. Given its large size and relative proximity, it has been observed over the entire wavelength range, from radio to gamma-rays. First discovered in the radio, in X-rays it manifests itself as a centrally-filled SNR. The X-ray morphology is dominated by a central bar with enhancements along the eastern and western edges. Recombining plasma has been identified along the central and western regions, with the plasma being more in  ionization equilibrium towards the eastern rim \citep{miceli2010, lopezetal13a, zv18}. The X-ray observations reveal that Si, S, Ar, Ca, and Fe exceed solar values, with Fe particularly concentrated in the eastern hemisphere \citep{hwang2000, miceli2006,lopezetal13b, zv18}. Emission lines from Cr and Mn have also been detected \citep{hwang2000, miceli2006}. 

W49B has been detected as a luminous source of gamma-ray emission using the LAT instrument aboard the Fermi $\gamma$-ray satellite \citep{abdo2010}. The observed spectrum from 0.2 - 200 GeV is consistent with either pion decay due to interaction with a dense medium, or with electron bremsstrahlung, with a high luminosity  in either case. It is inconsistent with a pulsar origin, thus suggesting that the $\gamma$-ray emission arises from shocked SN material. 

 \citet{R06}, in their survey of SNRs observed with Spitzer, noted that W49B showed [FeII] loops/filaments with H$_2$ emission on the edges, indicating ionic shocks within a molecular environment, predominantly on the E/SE side. \citet{K07} explored the infrared morphology in more detail. They found that the [FeII] emission appeared as co-axial rings. Infrared observations appear to suggest that the remnant is coincident with a surrounding partial shell of shocked H$_2$ \citep{R06, K07, Z14}. This has been interpreted as a sign that  W49B lies within a bubble surrounded by molecular clouds, with which it is interacting. The latter is also consistent with the pion decay interpretation of the gamma-ray emission.
 
 Currently, no compact object has been associated with the remnant and no pulsations have been detected \citep{gorham1996}, although a compact object cannot be absolutely ruled out \citep{zv18}.

\section{Application of SPI to W49B}
In order to derive the fundamental parameters describing W49B, and shed further light on its abundance measurements, we apply Smoothed Particle Inference (SPI) techniques to \textit{XMM-Newton} observations of W49B. Our aim is to thoroughly investigate the abundance of various elements and compare it to a wide range of SN explosion models, including asymmetric variants of core-collapse and Type Ia explosions, as well as energetic core-collapse models. The SPI technique takes advantage of the spatial and spectral information available in the \textit{XMM-Newton} data, to model the emission as a collection of smoothed particles (or gas blobs) and fit the superposed emission to the observation \citep{spi}. This method allows us to model the entire remnant and provides significant flexibility while analyzing the emission. Details of the SPI process, combined with a first demonstration of its unique capabilities,  were provided in two earlier papers, where it was used to study the SNR DEM L71 \citep{franketal19, Siegel2020}.

\subsection{SPI Model}
\label{section:spimodel}

\begin{table}
\caption{\textit{XMM}-Epic Observation 0724270101} 
\label{table:xmmobs}
\begin{tabular}{lccc}
\hline
\hline
Detector & Events & Net Exposure (ks)\\
 \hline 
MOS1 &424623&115\\
MOS2 &847239&115\\
pn &1699353&115\\
\hline
\end{tabular}
\end{table}

\begin{figure}[tbp]
\begin{center}
\includegraphics[width=\columnwidth]{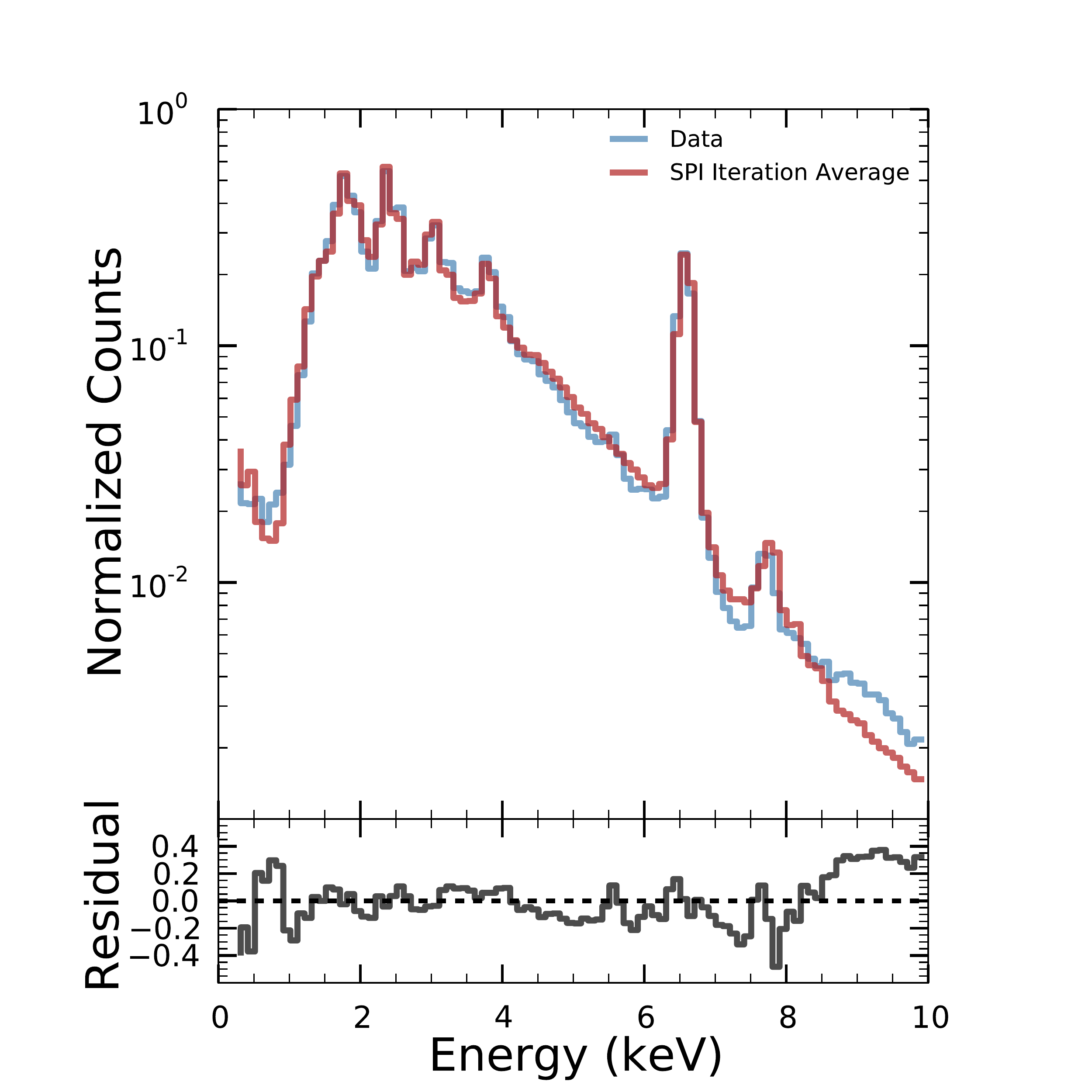}
\caption{The observed spectrum (blue) compared with the model spectrum (red) and the residual between the two; the residual is taken as the difference between the two divided by the observed data.} 
\label{fig:spectr}
\end{center}
\end{figure}

The SPI technique was applied to XMM EPIC observation 0724270101 of W49B from April 2014 (Table \ref{table:xmmobs}). A detailed description of the implementation is given in \citet{franketal19}. We used 300 isothermal gas blobs to fit the observed data. The number of blobs was chosen based on the spatial and spectral complexity of the SNR and to ensure each blob had $\sim10^4$ photons \citep{Frank2013}. Each blob was modeled by an independent \texttt{vmekal} model with \texttt{phabs} absorption to fit the SNR emission, along with several components to account for the different types of X-ray background emission \citep[see][for details]{franketal19}. The temperature, normalization, column-absorption density, and abundances of Si, S, Ar, Ca, Fe, and Ni were free and independent for each blob. Additionally, each blob had its own unique spatial model, with position and size treated as free parameters. Since multiple blobs can occupy the same line of sight, SPI has the flexibility to model multiphase plasma. Moreover, SPI does not assume an effective filling factor between temperature components along a line of sight, because SPI treats blob size as a free parameter. SPI also assumes no particular morphology or symmetry, which allows SPI to reproduce any arbitrary spatial distribution.
An MCMC algorithm was used to find the best set of model parameters for each iteration of the fitting process. The blobs from all converged iterations were used in the final analysis. 

The SPI fit of W49B had a reduced $\chi^2=1.2$. To further evaluate the quality of fit, we compare the model spectrum with the observed spectrum in Figure \ref{fig:spectr}. We find the model reproduces the major spectral features and that the two are in general agreement.

While overionized plasma has been detected in W49B, the current implementation of SPI is  incompatible with recent versions of \texttt{XSPEC}, which limits our model selection such that we are unable to incorporate newer models such as \texttt{vrnei}. This limitation comes into play when constraining the abundance of Ni in the western portion of the remnant. In order to better understand the effect on our abundance determination, we extract spectra from several regions within the central bar, where signatures of overionized plasma have been detected in past studies \citep{lopezetal13a, zv18}. We fit the extracted spectra in two different ways - with an absorbed thermal (\texttt{vmekal}) model with two temperature components; as well as an absorbed \texttt{vrnei} model with a cool \texttt{apec} component. We find that the two models return similar results for certain parameters. In the case of the abundance, the ionization equilibrium model returns marginally lower values for every element. However, for Si and Fe the effect is within the uncertainties, and for the remaining species the effect is not significant enough to impact our results. We additionally apply the same procedure to the western shell. In this region we find the Fe uncertainty to marginally increase, while Ni decreases, but is very poorly constrained by the collisional ionization equilibrium model. Our findings are consistent with those of \cite{ozawaetal09}, who find a marginal increase in Fe abundances and a significant decrease in Ni when the radiative recombination continuum is taken into account. The combination of their results and  our tests suggests that the effects from not taking the overionized plasma into account would lie within the stated uncertainties for all parameters except for Ni. Some prior studies \citep{Sato2020} have used the \cite{ozawaetal09} results to scale abundance measurements from CIE plasma to those expected in an overionized plasma. We do not implement this, and simply caution that our Ni value has a large uncertainty, and therefore cannot be used to discriminate between various abundance models.

\subsection{Low Temperature Component}
\label{sec:low_temp}
We use the \textit{XMM-Newton} response to estimate the number of emission lines detected as a function of the gas temperature and absorption for W49B. Based on this, we find a sub-population of blobs with high absorption and low kT for which too few emission lines would have been detected to accurately constrain the gas temperature and other properties. The kT distribution of this population, highlighted in Figure \ref{fig:hist_all}, is consistent with a $\sim$0.3 keV component. Prior studies have identified such a component and associated it with cool ISM material \citep{Kawasaki2005, zv18}. Unfortunately, due to the high absorption and low temperature, SPI does not have sufficient information to clearly detect this plasma or constrain its properties. Consequently, we choose to filter these blobs out at the start, and confine the remainder of our study to the warm(er) component. We note that if this potential swept-up ISM component is included, its total mass would be $\sim$30 $M_{\odot}$, for an assumed solar composition (see \S \ref{sec:mass_den} for details on mass estimation). Since prior studies have included a similar component in their analysis, we discuss this component when comparing to past mass estimates (see \S \ref{sec:mass_den} for discussion). However, we stress that this component is only ``potential" because the high absorption and low kT limits our ability to constrain the plasma.

Since one of our main goals is to investigate the progenitor and origin of W49B, the warm component is of significantly more interest, assuming that it contains the ejected plasma from the progenitor. For the rest of this paper, unless otherwise stated, we will be referring to this warmer component.

\subsection{Summary Results} 
\label{sec:summ}
With an assumed distance of 8 kpc, we find that the total volume emission measure (EM) of the warm component, obtained from the normalization of each blob, is EM = 1.5 ($\pm$0.1) $\times$ $10^{59}$ cm$^{-3}$. The EM-weighted abundances within the plasma, obtained using the filtered ensemble of blobs, are shown in Table \ref{table:emstats_angr}. We include the median and standard deviation ($\sigma$) of each posterior distribution, as well as the priors used in the SPI fitting. The median abundance values of all elements exceed solar, but with large standard deviations, indicating that the blobs populate a wide range of parameter space. 

\begin{figure*}
\gridline{\fig{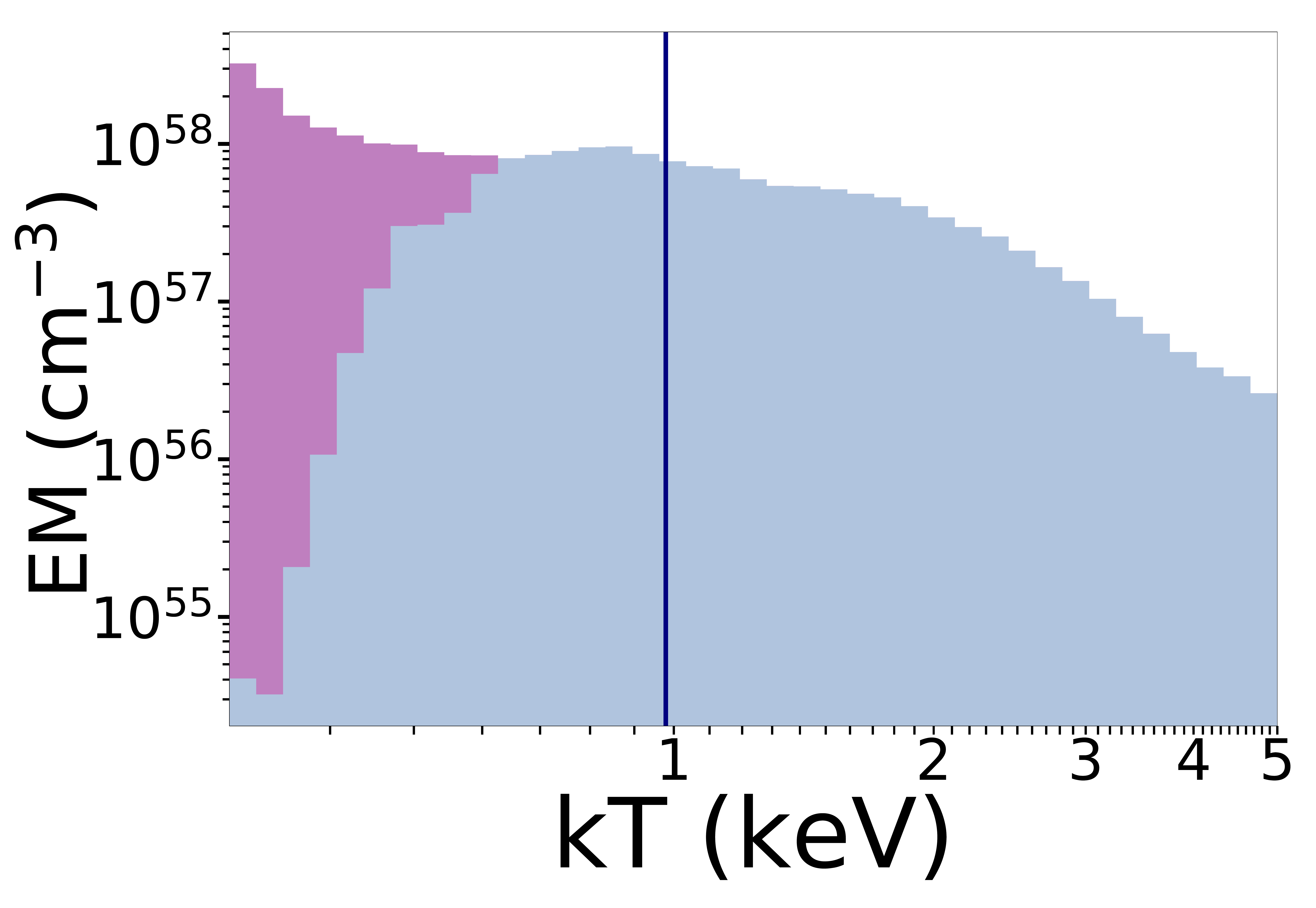}{0.33\textwidth}{} \fig{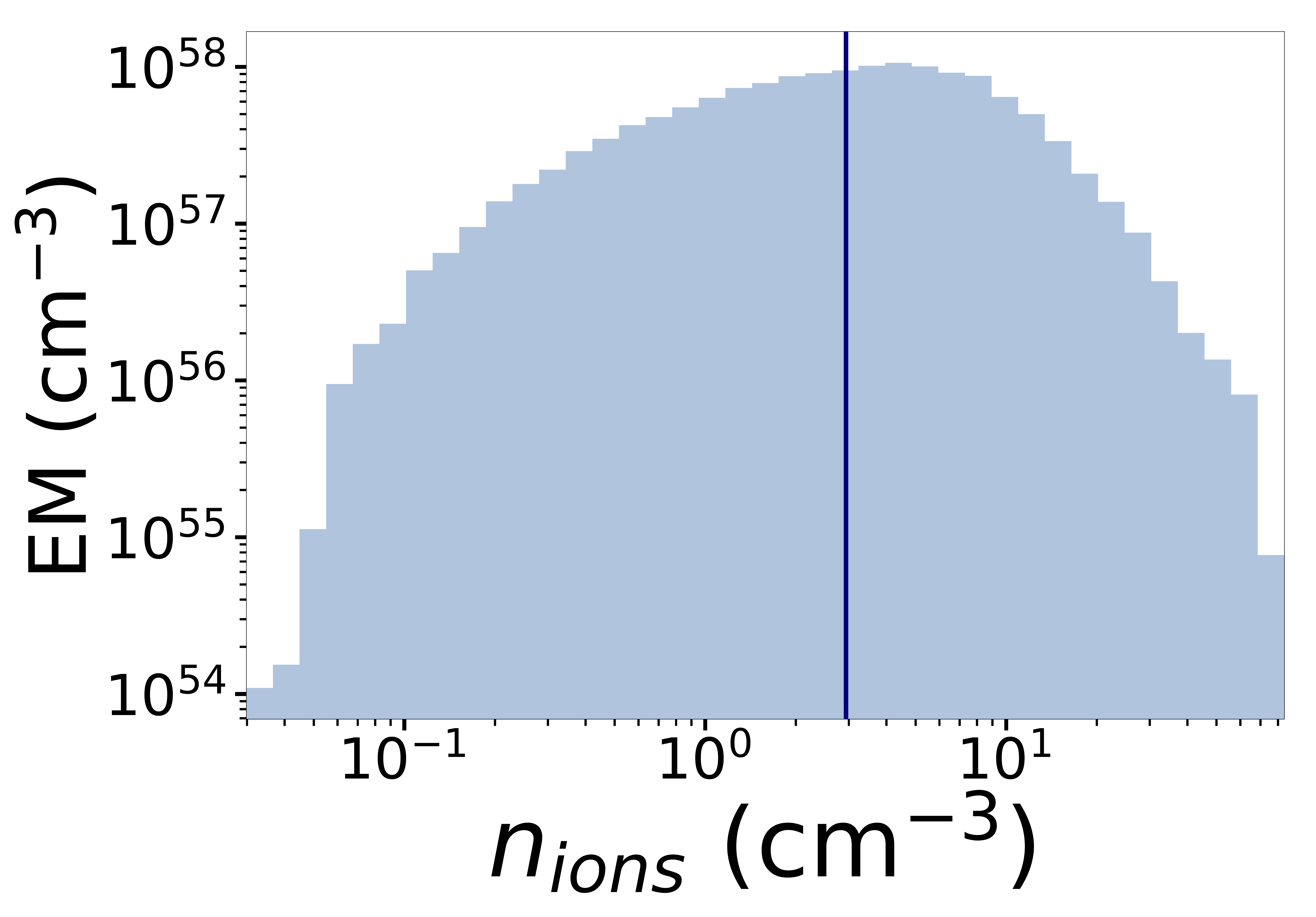}{0.33\textwidth}{} \fig{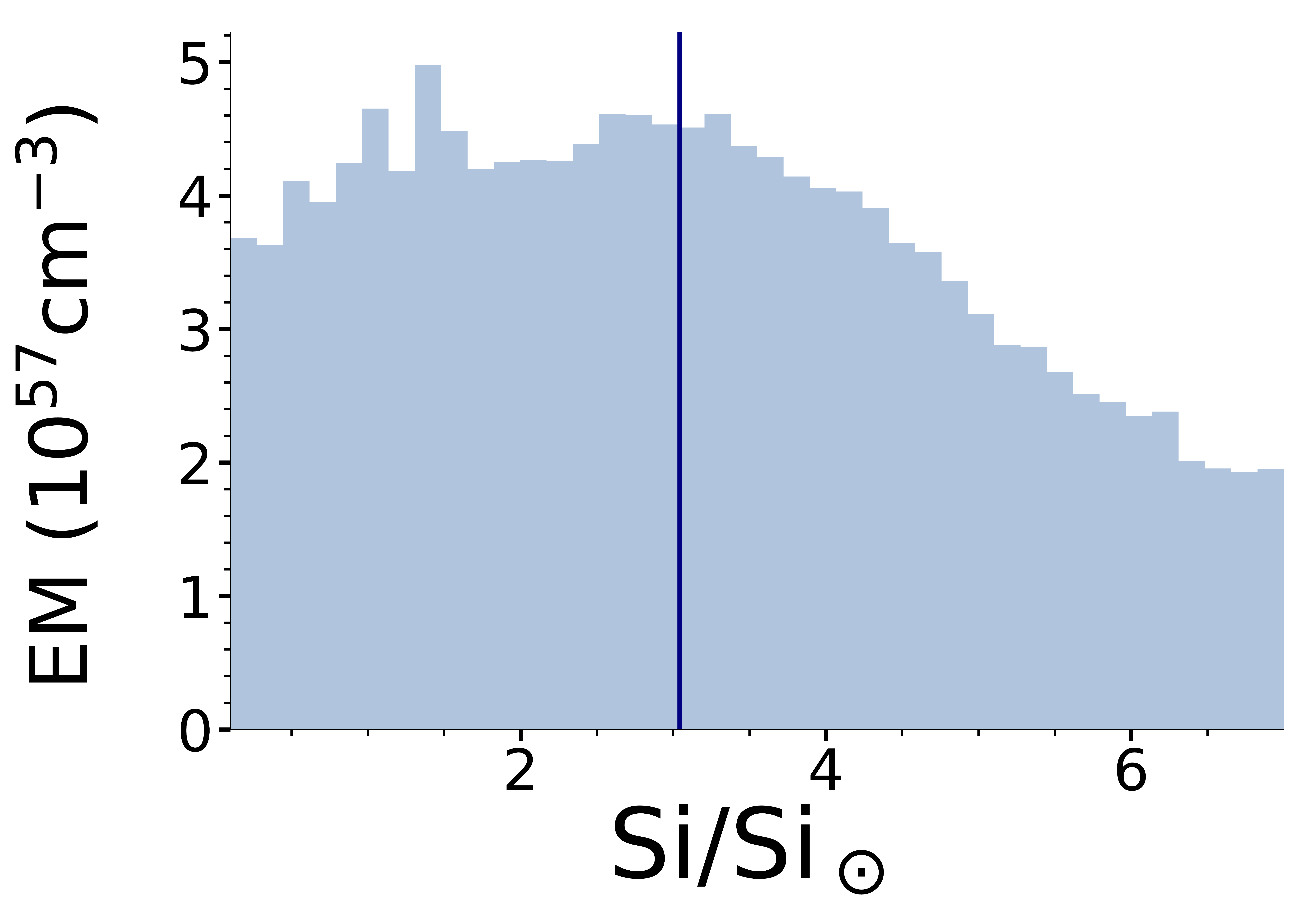}{0.33\textwidth}{}}
\gridline{\fig{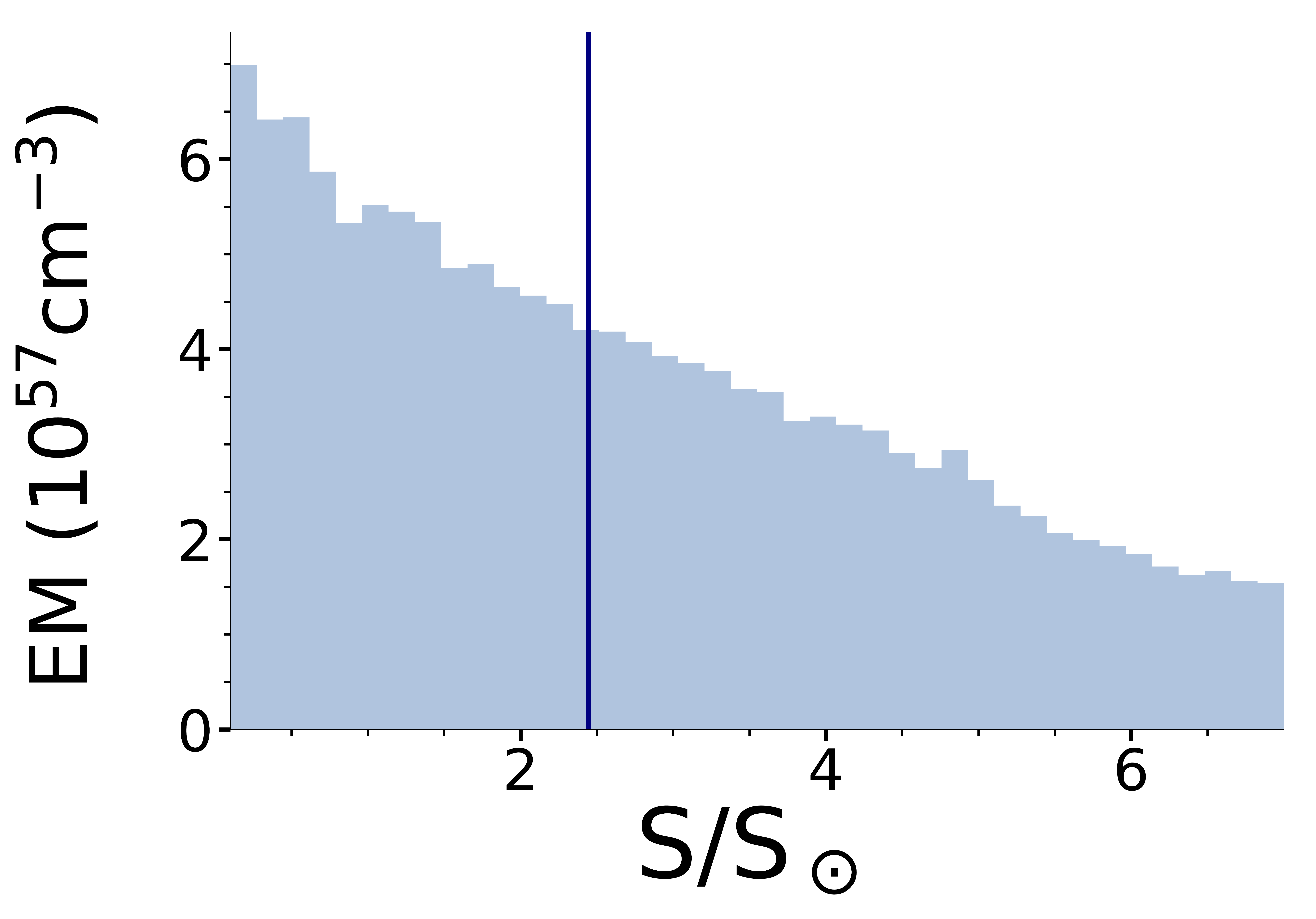}{0.33\textwidth}{} \fig{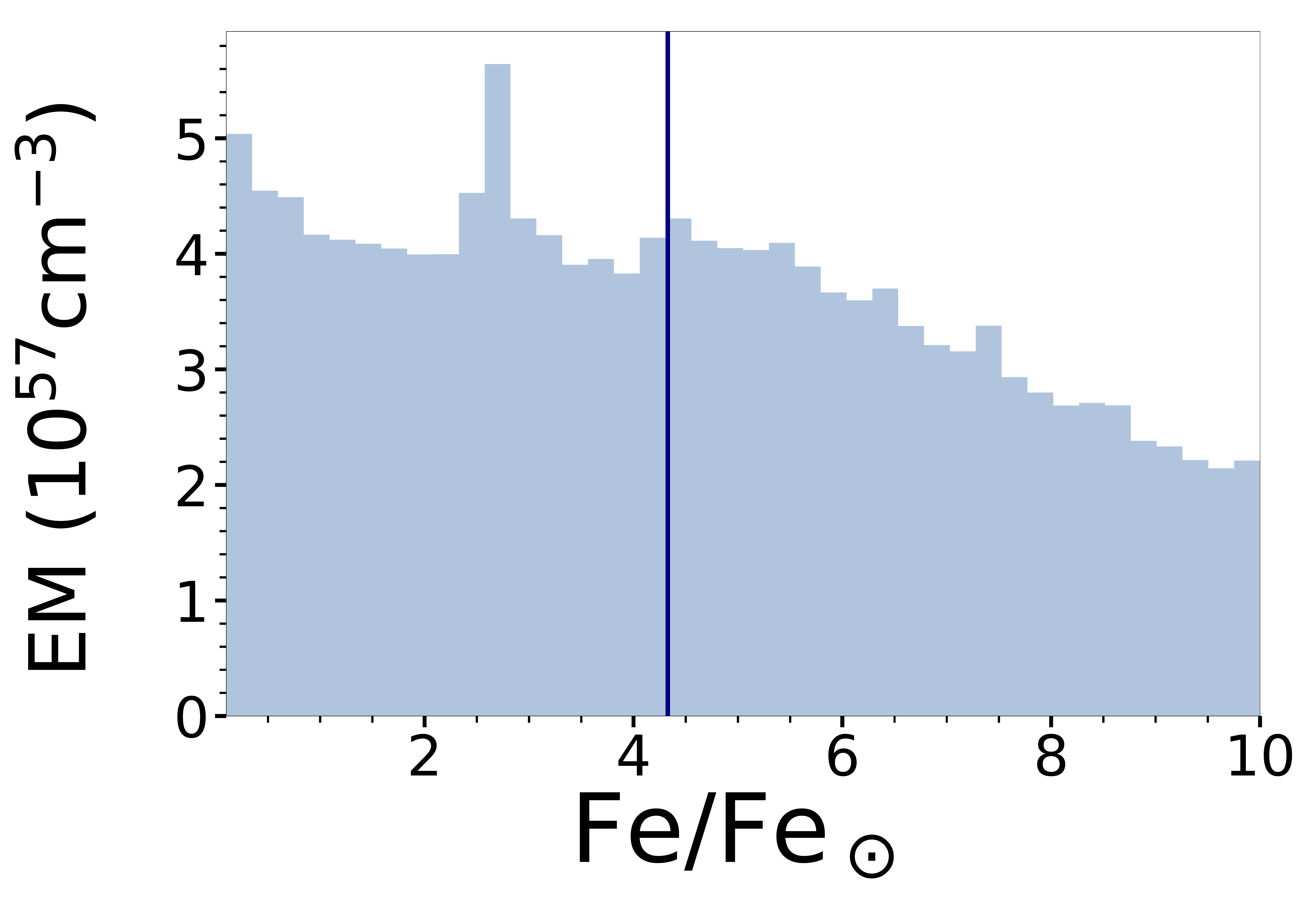}{0.33\textwidth}{}
}
\caption{Emission-measure-weighted parameter posterior distributions. Units of abundance are relative to solar values in \citet{angr}. kT is in units of keV, and emission measure is in units of cm$^{-3}$ or $10^{57}$ cm$^{-3}$. The median is shown as a solid line. For the kT distribution, the cool ISM component is highlighted in purple.
}
\label{fig:hist_all}
\end{figure*}

To further evaluate the properties of the plasma, we produce EM-weighted posterior distributions for kT and ion density (see \S \ref{sec:mass_den}). For the kT distribution, we also include the low temperature component discussed in \S \ref{sec:low_temp}. These plots are shown in Figure \ref{fig:hist_all}, alongside those of Si, S, and Fe. The distributions reveal the wide range for each parameter. In particular, Si, S, and Fe all show considerable emission from blobs with super-solar abundances, while blobs with a density up to $\approx$70 cm$^{-3}$ are also seen to be present. We find that while prior studies reported a maximum kT of $\sim$2.5 keV, the temperature distribution of our blobs actually extends past $\sim5$ keV. We attribute this extended range to the ability of SPI to characterize all components from all parts of the remnant, including those with a lower emission measure. Previous studies have generally used two temperature components for each spatial bin, which results in two dominant temperatures, but fails to capture the full range. We note that the highest EM portions of the temperature distribution in our study are consistent with the overall range given in previous studies.

\begin{table}
\begin{center}
\caption{Priors and EM-weighted Summary Statistics} 
\label{table:emstats_angr}
\begin{tabular}{lccc}
\hline
\hline
& Prior & Median & $\sigma$\\
\hline
$N_{H}$ ($10^{22}$ cm$^{-2}$) &1.0-10.0& 7.0 $\pm$ 0.4 & 2.3 $\pm$ 0.1 \\
kT (keV) &0.1-5.0& 0.98 $\pm$ 0.06 & 0.68 $\pm$ 0.02 \\
Si/Si$_{\odot}$ &0.1-7.0& 3.04 $\pm$ 0.28 & 1.83 $\pm$ 0.09 \\
S/S$_{\odot}$ &0.1-7.0& 2.45 $\pm$ 0.28 & 1.86 $\pm$ 0.08 \\
Ar/Ar$_{\odot}$ &0.1-7.0& 3.05 $\pm$ 0.24 & 1.79 $\pm$ 0.09 \\
Ca/Ca$_{\odot}$ &0.1-8.0& 4.44 $\pm$ 0.32 & 2.06 $\pm$ 0.12 \\
Fe/Fe$_{\odot}$ &0.1-10.0& 4.33 $\pm$ 0.46 & 2.72 $\pm$ 0.15 \\
Ni/Ni$_{\odot}$ &0.1-10.0& 5.11 $\pm$ 0.42 & 2.63 $\pm$ 0.14 \\
\hline
\end{tabular}
\end{center}
\tablecomments{Abundances are relative to solar, as defined by \citet{angr}. Temperature prior is uniform in log kT.}
\end{table}

\begin{figure*}
\gridline{\fig{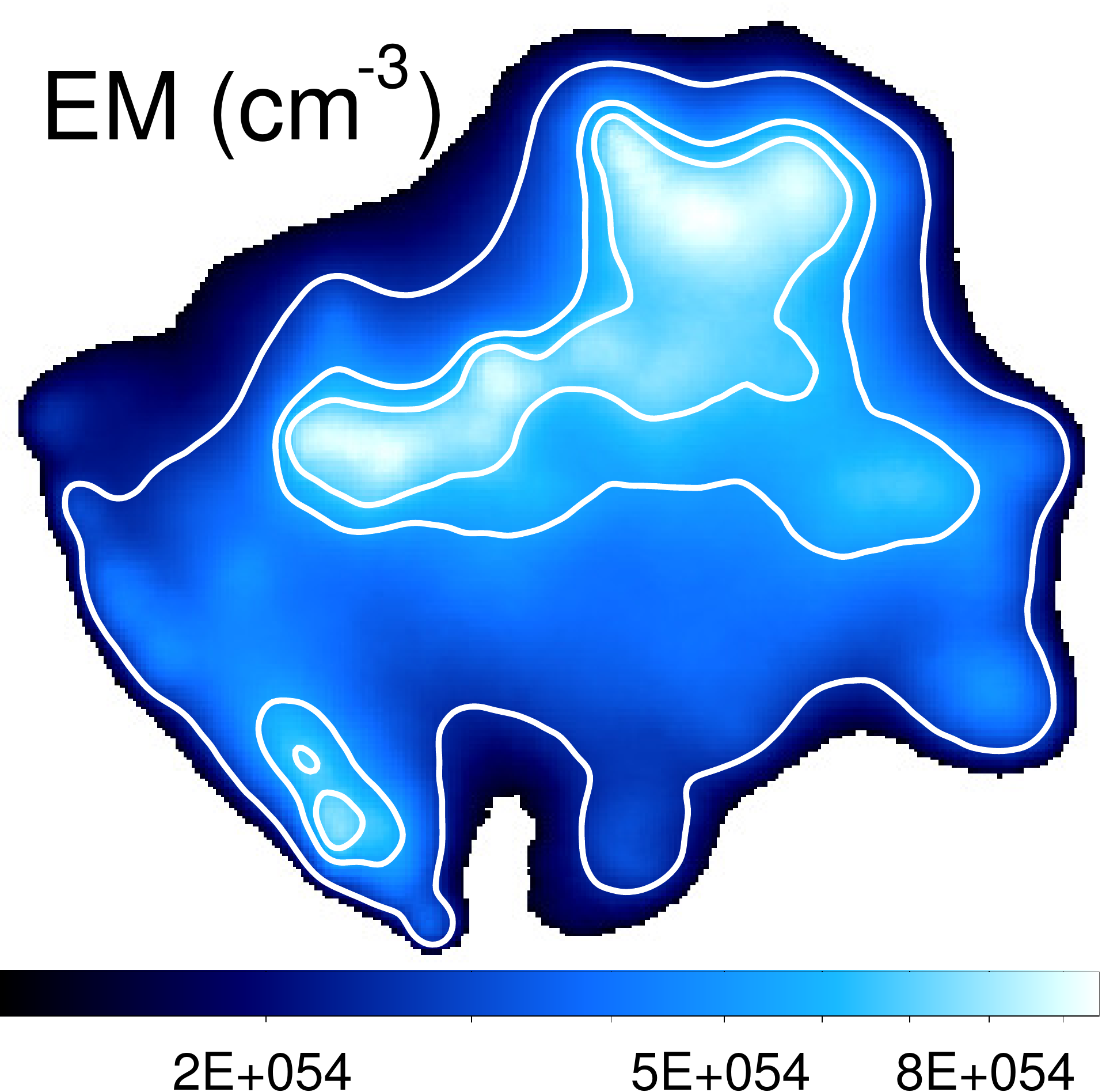}{0.33\textwidth}{} \fig{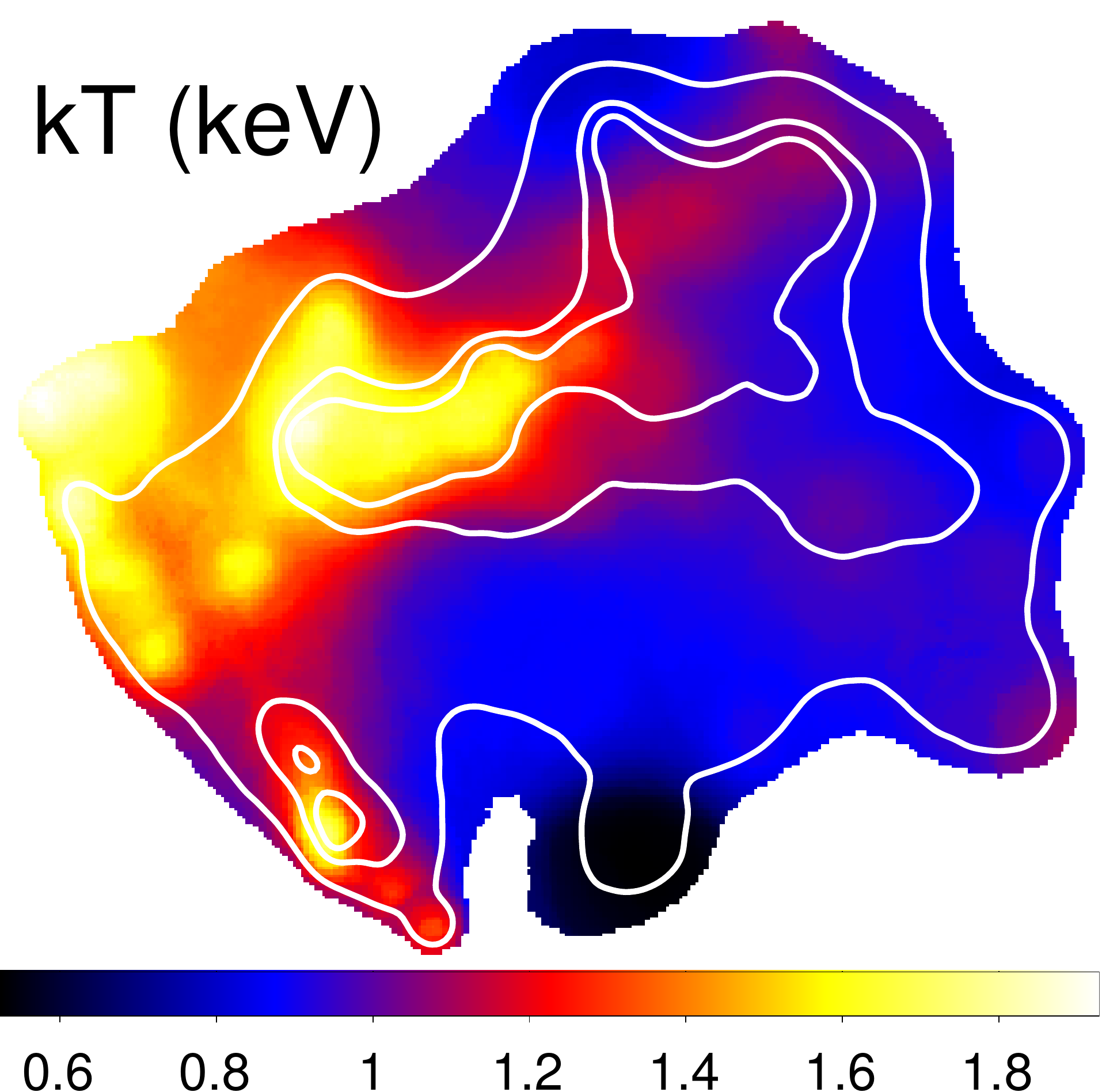}{0.33\textwidth}{} \fig{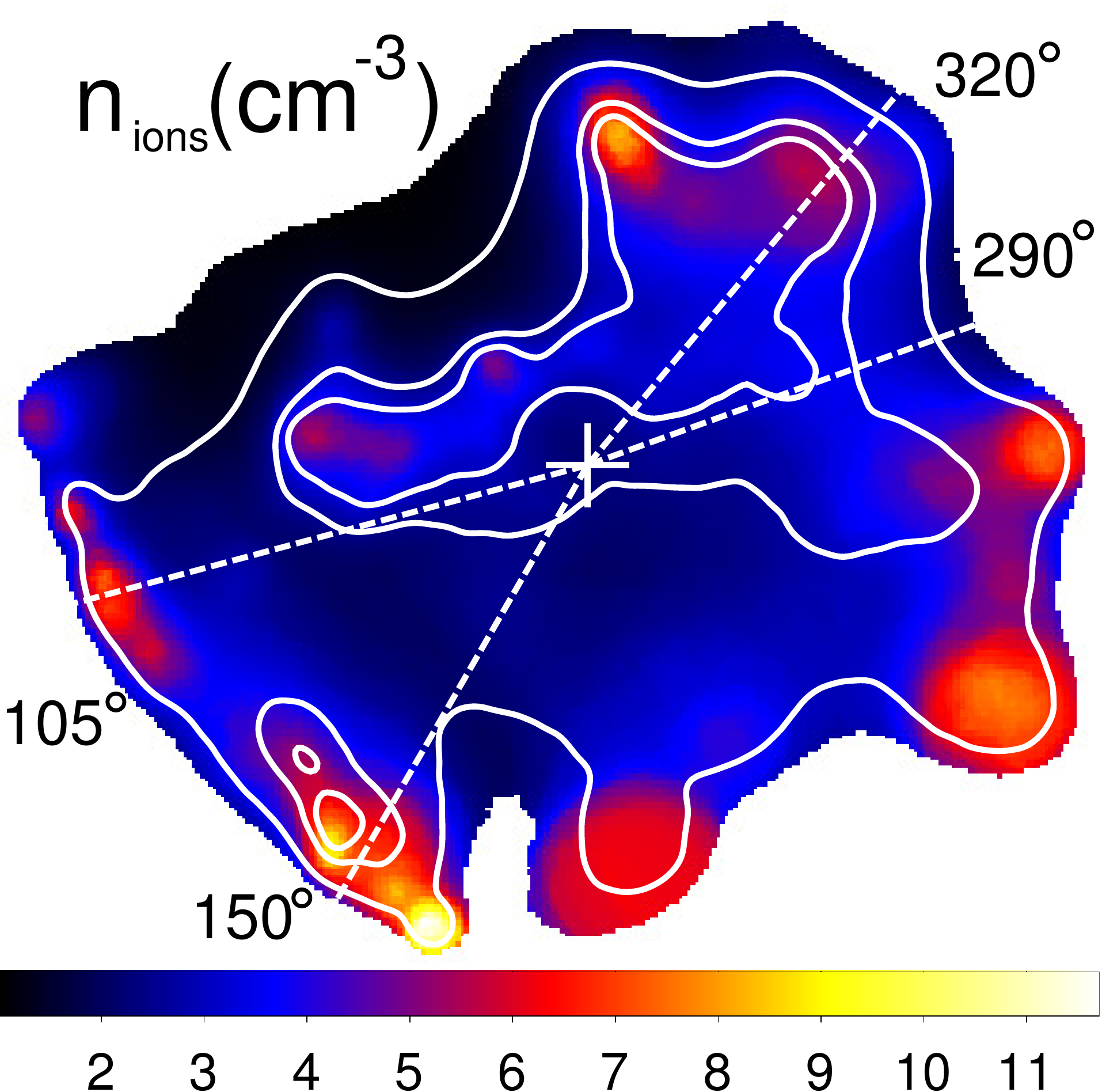}{0.33\textwidth}{}}
\gridline{\fig{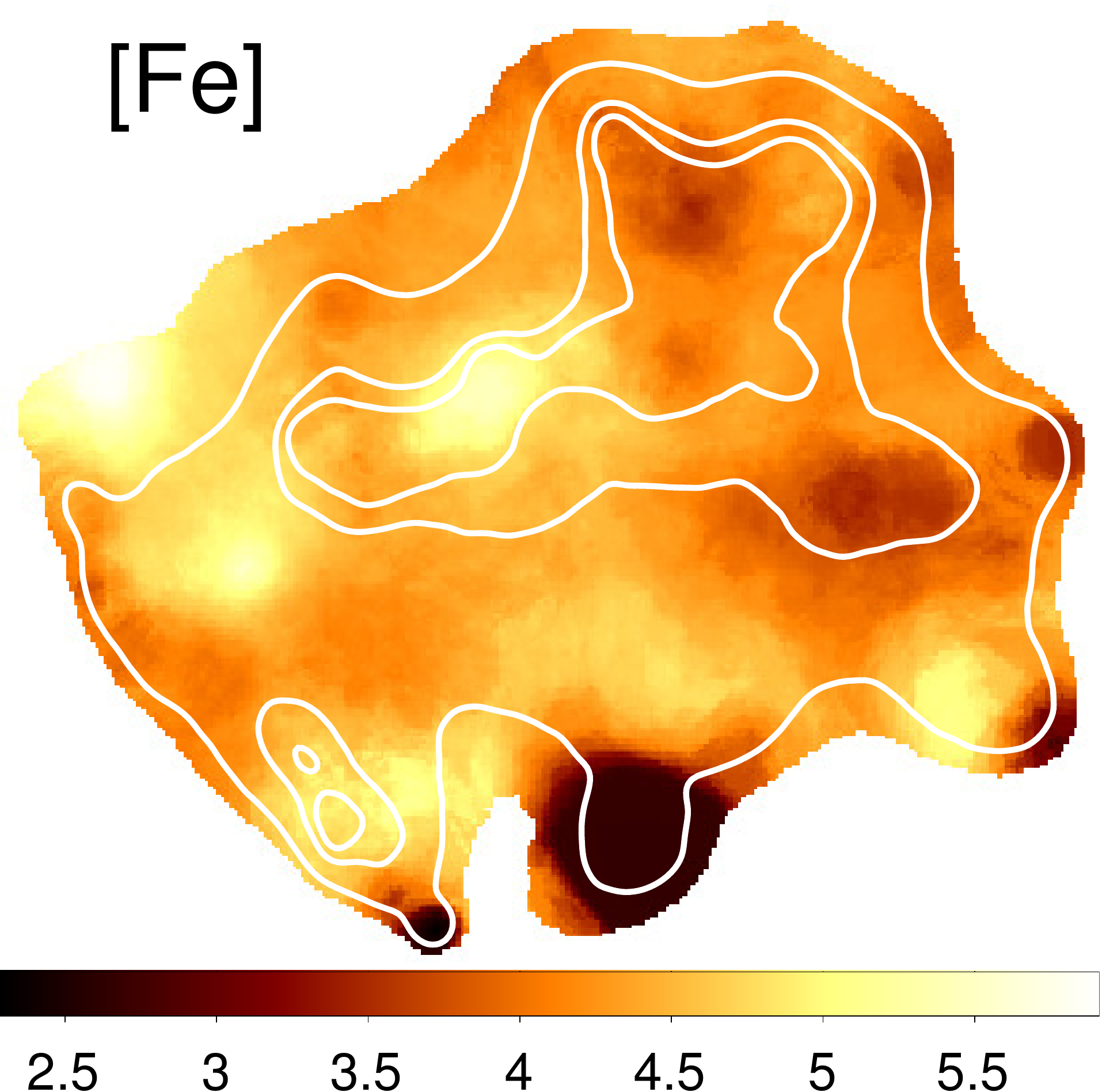}{0.31\textwidth}{} \fig{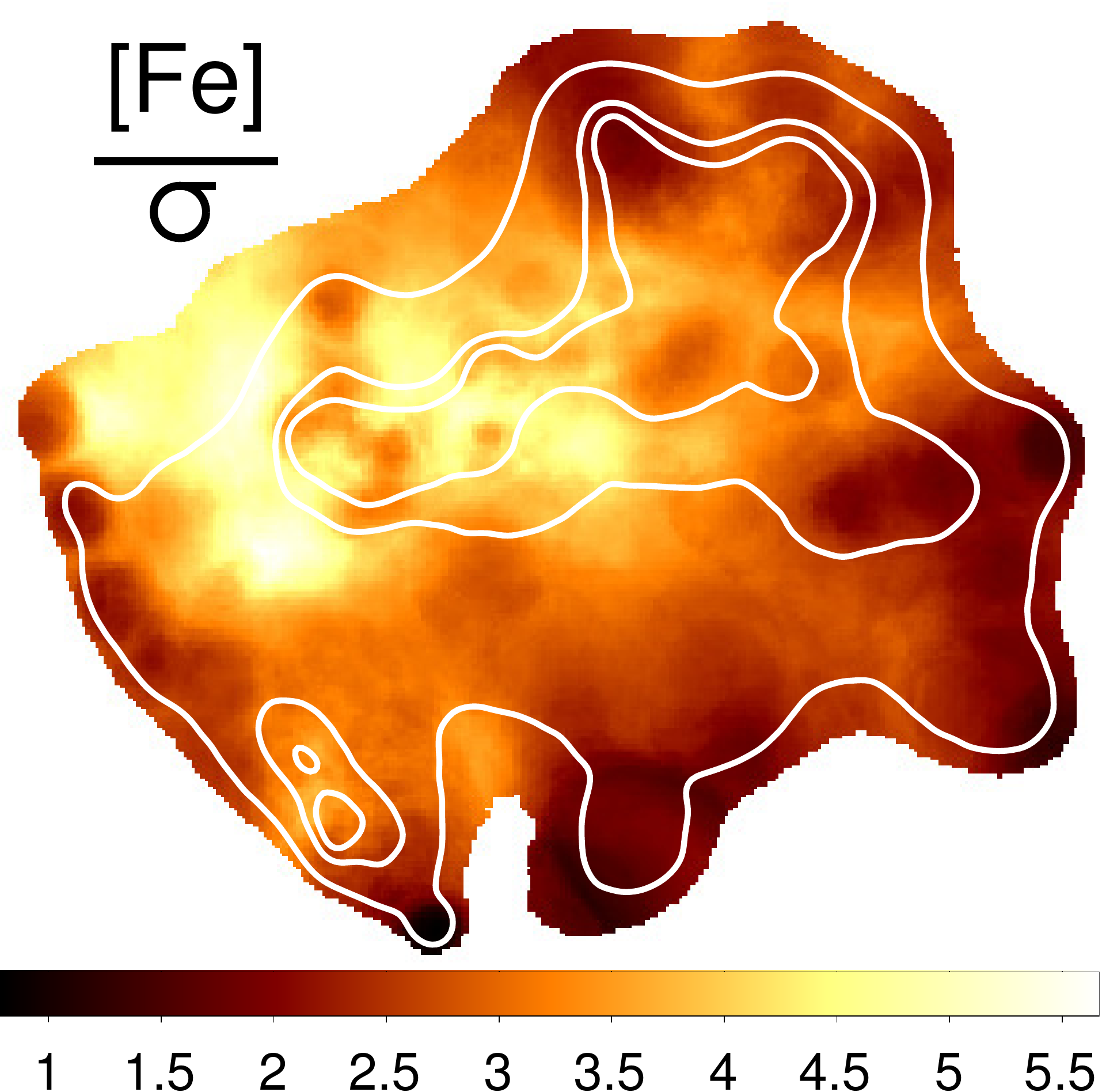}{0.31\textwidth}{}}
\caption{\textit{Top Row} from left to right: Maps of EM (cm$^{-3}$, square root scale), EM-weighted temperature (keV), EM-weighted ion density (cm$^{-3}$). The EM map is produced by taking the total value within each spatial bin, and the other maps are produced by taking the median in each spatial bin. \textit{Bottom Row} from left to right: Maps of EM-weighted median Fe relative to solar from \citet{angr} and the EM-weighted median Fe divided by the one sigma error. The white contours are taken from the EM map and for each map the values are set to zero where the EM was lower that 0.05\% of the mean to avoid noise due to poor statistics on the outer edges.
}
\label{fig:general_maps}
\end{figure*}

Using the location and size of each blob, we create maps for all parameters. Maps are created by dividing the SNR region into spatial bins of 1 arcsec. In each spatial bin, the contributions from blobs that overlap that bin are combined to assign a single value to a given parameter, as in \citet{franketal19}. In Figure \ref{fig:general_maps}, the emission measure map is shown, accompanied by the EM-weighted maps of the temperature, ion density, and Fe. In order to account for the uncertainty in the abundance values, we also provide a significance map for Fe, defined as the EM-weighted median Fe abundance divided by the one sigma error. We find large spatial variations among nearly all parameters, in particular a strong southwest to northeast temperature gradient. Interestingly, Fe appears to be widely distributed over the entire remnant rather than concentrated in a small region. Due to the EM weighting, only the high emission measure components are reflected in the maps; this results in a smaller range for the parameters in the maps in Figure \ref{fig:general_maps} than the histograms in Figure \ref{fig:hist_all}.

\begin{table}
\begin{center}
\caption{Individual Element Mass Yields} 
\label{table:yields}
\begin{tabular}{lccc}
\hline
\hline
 & SPI Value & Solar Value & Difference \\
 \hline 
Si (\msun) & 0.30 $\pm$ 0.04 & 0.09 $\pm$ 0.01 & 0.21 $\pm$ 0.04 \\
S (\msun) & 0.16 $\pm$ 0.02 & 0.05 $\pm$ 0.01 & 0.11 $\pm$ 0.02 \\
Ca (\msun) & 0.04 $\pm$ 0.01 & 0.01 $\pm$ 1E-3 & 0.03 $\pm$ 0.01 \\
Ar (\msun) & 0.04 $\pm$ 0.01 & 0.01 $\pm$ 2E-3 & 0.03 $\pm$ 0.01 \\
Fe (\msun) & 0.97 $\pm$ 0.17 & 0.24 $\pm$ 0.03 & 0.73 $\pm$ 0.17 \\
Ni (\msun) & 0.05 $\pm$ 0.01 & 0.01 $\pm$ 1E-3 & 0.04 $\pm$ 0.01 \\
Total (\msun) & 1.56 $\pm$ 0.18 & 0.41 $\pm$ 0.07 & 1.15 $\pm$ 0.18 \\
\hline
\end{tabular}
\end{center}
\end{table}

\subsection{Mass and Density}
\label{sec:mass_den}
Using EM $=\int n_e n_H dV$, the volume of each blob, and a constant density per blob, we are able to calculate the density throughout the remnant. Using the blob volume in the equation for EM above,  we derive the combination of $ n_e n_H$; we further need a relation between $n_e$ and $n_H$ to extricate the individual quantities. This process was discussed in \citet{Siegel2020}. Here we consider two different approaches.

In the first case, we assume that all the material has a solar composition. We then use the EM and volume from the fit, along with the $n_e/n_H$ ratio corresponding to solar values, to calculate the density within each blob. This nominally represents what the density would be if the gas were composed only of swept-up local Galactic material.

In the second approach, the abundance of each element is taken from the SPI fit. Any abundance greater than solar is likely associated with the ejecta component. Since for each element SPI assigns an independent abundance value to each blob, each blob will consequently have a different value of $\mu$, the mean molecular weight, and the ratio $n_e/n_H$. The density of each blob is calculated using these values in conjunction with the associated EM and volume.

Once the density of a blob has been calculated, the total blob mass ---the product of the blob density and volume--- and the mass of each individual species within the blob is calculated. The latter is obtained using the composition of the material to find the element's mass fraction. Once we have calculated the overall mass and individual species mass within each blob, the total mass of material comprising the remnant is found by summing over all blobs.

Using this second approach, the total mass calculated for W49B is M$_T$=130 $(\pm 16)$ \msun, with an EM-weighted median ion density of n$_{ions}$=2.9 ($\pm$ 0.3) cm$^{-3}$. We find that after taking the mass contribution from our potential cool ISM component ($\sim$30 {\msun}, see \S \ref{sec:low_temp}) into account, our inferred total mass for W49B lies in-between the nominal estimate ($\sim$308 {\msun}) and the thin-shell geometry estimate ($\sim$130 {\msun}) from \cite{zv18}, for our assumed distance of 8 kpc. We note though that we have adopted different thresholds for the ISM material. 

This total mass suggests that, irrespective of the origin and resulting ejecta mass, a large amount of mass has been swept-up by the SN shock wave. Comparing the mass of each individual element in both approaches provides a rough estimate of the contribution of the ejecta and the swept material. In Table \ref{table:yields}, the total mass of each element is given using either solar or SPI abundance values, and their difference. If we assume the only source of extra-solar values is the SN ejecta, then the difference between the two values is approximately the mass ejected in the SN explosion. We therefore calculate an ejecta mass of M$_{ej}$=1.2 $(\pm 0.2)$ \msun. If we merely consider the mass of all metals calculated using the SPI technique, assuming no contribution from the surrounding material, then M$_{ej}$=1.6 $(\pm 0.2)$ \msun. We note that, given the uncertainties in the calculation, both of these values are compatible with a Type Ia progenitor, of either Chandrasekhar or sub-Chandrasekhar mass. It should be emphasized that the calculation comes with many inherent caveats: the assumption that all deviations from solar abundance are due to ejecta material; the assumption that the neighborhood of W49B has a solar composition;
and possible deviations between ambient abundance values in different regions of W49B.  

In Figure \ref{fig:mass_distr}, we plot the normalized total mass, and the mass of Si, S, and Fe as a function of position angle (where 0$^{\circ}$ is to the North). We find that the mass distribution shows distinct peaks, with only one peak in the eastern hemisphere, but multiple on the western side. The  Si distribution closely follows the overall one. However, S and Fe seem somewhat depleted in the western hemisphere. The existence of the eastern peak is somewhat consistent with the assumption of a jet-like structure in that region, as suggested by \citet{lopezetal09}, but the presence of multiple peaks on the western side disagrees with this line of reasoning. At the very least, there do not exist equal and opposite jets, and the distribution on the western side is inconsistent with a jet-like structure.

\section{Discussion}
\label{sec:Disc}
We use the morphology and abundance information provided by SPI to examine the interaction between the remnant and the surrounding medium and to compare the abundance values with a selection of SN models. Using these results we explore the origin of W49B.

\subsection{Interaction with the Surrounding Medium}

\begin{figure}[htbp]
\includegraphics[width=\columnwidth]{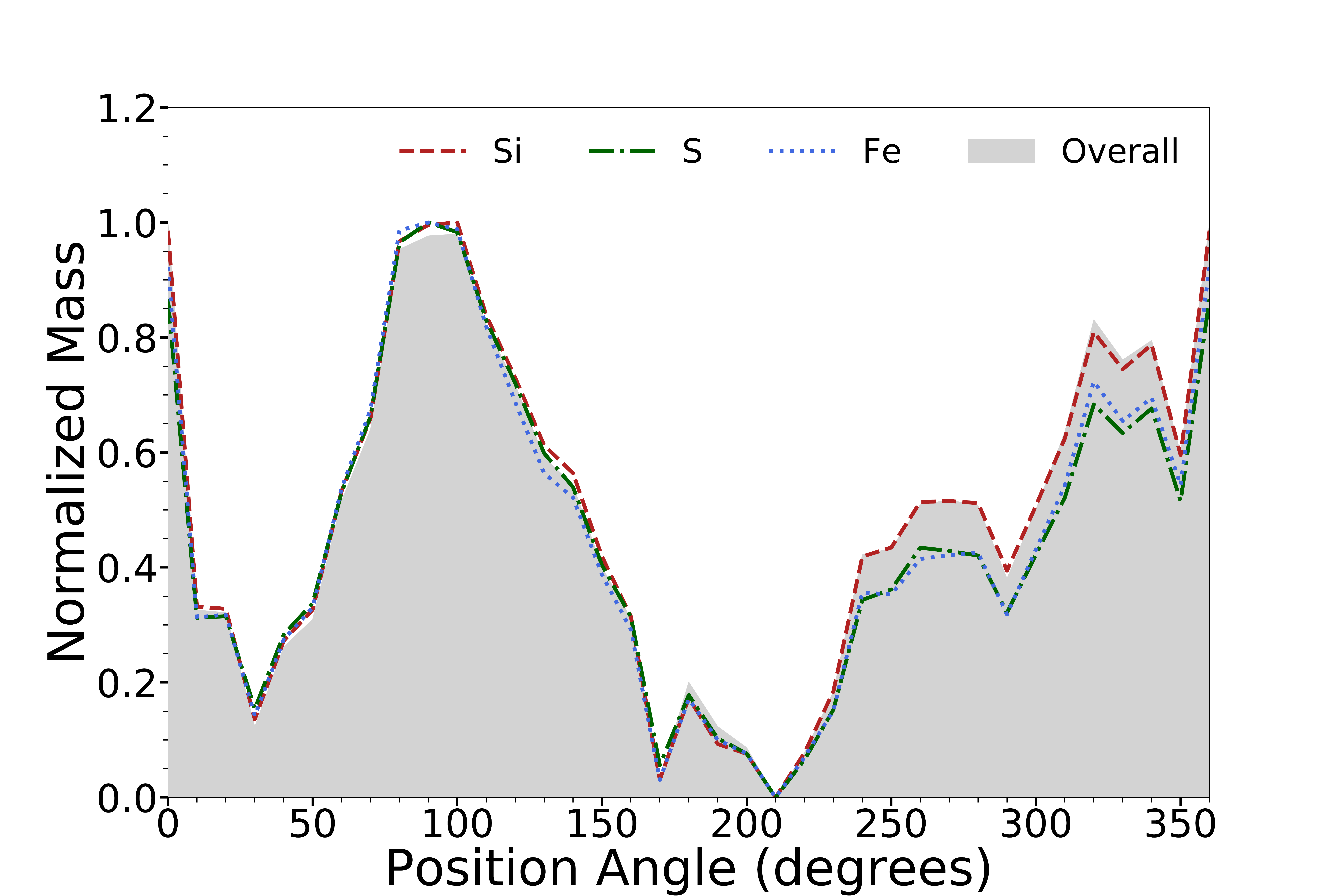}
\caption{Total mass and the individual element mass of Si, S, and Fe normalized and plotted against position angle. The geometric center is shown on the median density map (Figure \ref{fig:general_maps}, upper right).} 
\label{fig:mass_distr}
\end{figure}

\label{sec:surr}
\begin{figure}[htbp]
\includegraphics[width=\columnwidth]{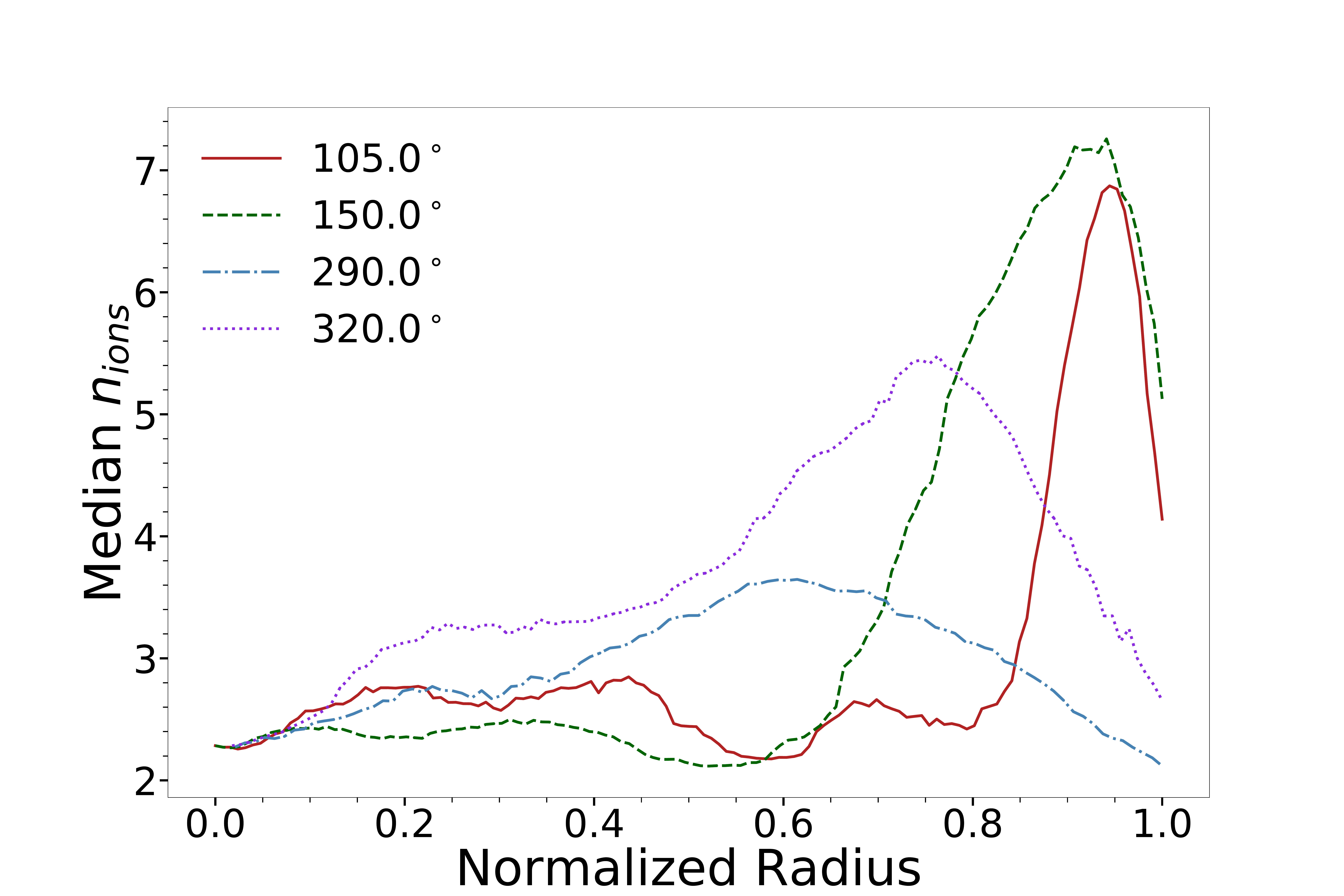}
\caption{Profiles of EM-weighted median ion density against normalized radius for a selection of position angles in units of cm$^{-3}$. The geometric center and position angles are shown on the median density map (Figure \ref{fig:general_maps}, upper right).} 

\label{fig:profiles}
\end{figure}

As discussed in \S \ref{sec:mw}, studies of W49B in the near-infrared have revealed the presence of [Fe II] emission along coaxial rings, and a shell of shocked molecular hydrogen with enhancements along the eastern edge, suggesting the remnant is interacting with dense gas predominantly towards the E/SE \citep{R06, K07, Z14}. The gamma-ray data from Fermi \citep{abdo2010} is consistent with this interpretation. Moreover, \citet{lopezetal13a} finds that the plasma on the eastern side is in collisional ionization equilibrium, whereas that to the west and center is recombining. This suggests a higher density on the eastern side to allow ionization equilibrium to be reached in a time-span of a few thousand years, consistent with the dense gas interpretation.

The densities inferred with SPI are consistent with this picture. The EM-weighted median density map (Figure \ref{fig:general_maps}, upper right), shows a slight enhancement along the horizontal bar and western shell, with the highest enhancement appearing along the eastern edge.

To investigate this aspect further, we plot density profiles for various position angles (Figure \ref{fig:profiles}). The selected position angles were chosen to represent almost diagonally opposite regions on both western and eastern edges. The geometric center and selected position angles are shown on the median density map (Figure \ref{fig:general_maps}, upper right).

The profiles show that the eastern edge has a higher maximum value of the density and a narrower peak than the west. This behavior is in agreement with the  scenario of W49B interacting with a higher density region along the eastern edge. We note that the density along the eastern edge could be substantially higher than this in some locations, with densities reaching 70 cm$^{-3}$ as evidenced from the density distribution (Figure \ref{fig:hist_all}), but those regions are few and harder to delineate.

\subsection{Abundance Comparison to SN Models}
To infer the nature of W49B's progenitor star, we compare the abundances calculated from SPI with SN explosion models. Since core-collapse, Type Ia, and hypernova models have been used in the past to model the abundance pattern, we consider a large suite of explosion models that take all of these mechanisms into account \citep{mn03, nomotoetal06, maedaetal10, seitenzahletal13, Fink2014, sukhboldetal16, ln18, Limongi2018, bravo2019, Leung2020}. To enable efficient comparison, we divide the explosion models into 7 families, each one describing a different explosion mechanism. Descriptions and references for each model family are listed in Table \ref{tab:families}. After comparing the yields from all models to the abundances and masses of various elements 
as derived by SPI, we choose, for the sake of brevity, a representative sample of models (listed in Table \ref{table:models}) for further detailed comparison.

\begin{deluxetable*}{lccl}
\tablecaption{Summary of Model Families\label{tab:families}}
\tablehead{
\colhead{Family} & \colhead{SN Type} & \colhead{Number of Models} &
\colhead{References}
}
\startdata
Deflagration-to-Detonation Transition (DDT) & Ia & 139 & M10, S13, LN18, B19\\
Double-Detonation (DD) & Ia & 40 & LN20\\
Pure Deflagration (DEF) & Ia & 19 & M10, F14, LN18\\
Off-center Deflagration-to-Detonation Transition (ODDT) & Ia & 1 & M10\\
Standard Energy core-collapse (SCC) & CC & 44 & S16, L18\\
Hypernova (HN) & CC & 4 & N06\\
Bipolar Hypernova (BHN) & CC & 8 & M03\\
\enddata
\tablecomments{M03: \citet{mn03}, N06: \citet{nomotoetal06}, M10: \citet{maedaetal10},  S13: \citet{seitenzahletal13}, F14: \citet{Fink2014}, S16: \citet{sukhboldetal16}, L18: \citet{Limongi2018}, LN18: \citet{ln18}, B19: \citet{bravo2019}, LN20: \citet{Leung2020}}
\end{deluxetable*}

\begin{deluxetable*}{lllc}
\tablecaption{Selected SN Models for Comparison\label{table:models}}
\tablehead{
\colhead{Model Name\tablenotemark{a}} & \colhead{Family} & \colhead{Description} &
\colhead{References}
}

\startdata
40$M_{\odot}$ 0.02$Z_{\odot}$ 30$E_{51}$ & HN & Symmetric Hypernova & N06 \\
40A & BHN & 40M$_{\odot}$ Bipolar Hypernova $\mid$ $\theta_{jet}$=15$^{\circ}$ $\mid$ M$_{REM0}$\tablenotemark{b} $=1.5M_{\odot}$& M03 \\
40B & BHN & 40M$_{\odot}$ Bipolar Hypernova $\mid$ $\theta_{jet}$=45$^{\circ}$ $\mid$ M$_{REM0}=1.5M_{\odot}$& M03 \\
40D & BHN & 40M$_{\odot}$ Bipolar Hypernova $\mid$ $\theta_{jet}$=15$^{\circ}$ $\mid$ M$_{REM0}=3.0M_{\odot}$& M03 \\
040a000 & SCC & Non-rotating, solar metallicity, symmetric 40M$_{\odot}$ & L18 \\
W18 s12.5 & SCC & Symmetric 12.5M$_{\odot}$ & S16 \\
\hline
300-1-c3-1 & DDT & Benchmark density DDT model & LN18\\
O-DDT & ODDT & Off-center deflagration ignition & M10 \\
N100 & DDT & DDT with 100 ignition points & S13 \\
M$_{CH}$ $\xi=$0.9 $\rho_{DDT}$=2.8 $\times$ 10$^7$ $Z=$9.00 $\times$ 10$^{-3}$ \tablenotemark{c} & DDT & Symmetric M$_{CH}$ DDT with modified C/O reaction rate & B19 \\
300-1-c3-1P & DEF & Benchmark density deflagration model & LN18 \\
\hline
\enddata
\tablenotetext{a}{For clarity, the listed model names are unchanged from the corresponding reference paper}
\tablenotetext{b}{Mass of the central remnant (compact object)}
\tablenotemark{c}{$\rho_{DDT}$ is the central density and is given in g cm$^{-3}$}
\end{deluxetable*}

\citet[][hereafter L13b]{lopezetal13b} compared the abundance values derived from their X-ray fitting to a selection of core-collapse models \citep{mn03, nomotoetal06}, and concluded that those with an aspherical explosion provided the best fit. \citet[][hereafter ZV18]{zv18}, on the other hand, found that Type Ia models matched best after comparing their observations to a subset of both core-collapse and Type Ia models \citep{mn03, nomotoetal06, maedaetal10, seitenzahletal13, sukhboldetal16}. To enable comparison with prior work, we include the abundance values derived by L13b and ZV18 in the following analysis. 

L13b and ZV18 both analyze \textit{Chandra} observations, applying a binning procedure that breaks the remnant into small regions and fits the spectra separately in each region. For the spectral fitting, L13b employed the absorption component \texttt{phabs} multiplied by the thermal emission \texttt{vmekal} model in XSPEC, while ZV18 used the  \texttt{vapec} thermal emission model, or alternatively the \texttt{vrnei} radiative recombination model, with an additional cool component modeled by \texttt{apec}. Both studies adopted solar abundances cited by \cite{Asplund2009}, which we converted to those given by \citet{angr} for comparison. Thus, a comparison of our values (obtained using an {\it XMM-Newton} observation with the SPI method) with these two represents a comparison across a range of methods, instruments, and spectral models.

\begin{figure}[htbp]
\includegraphics[width=\columnwidth]{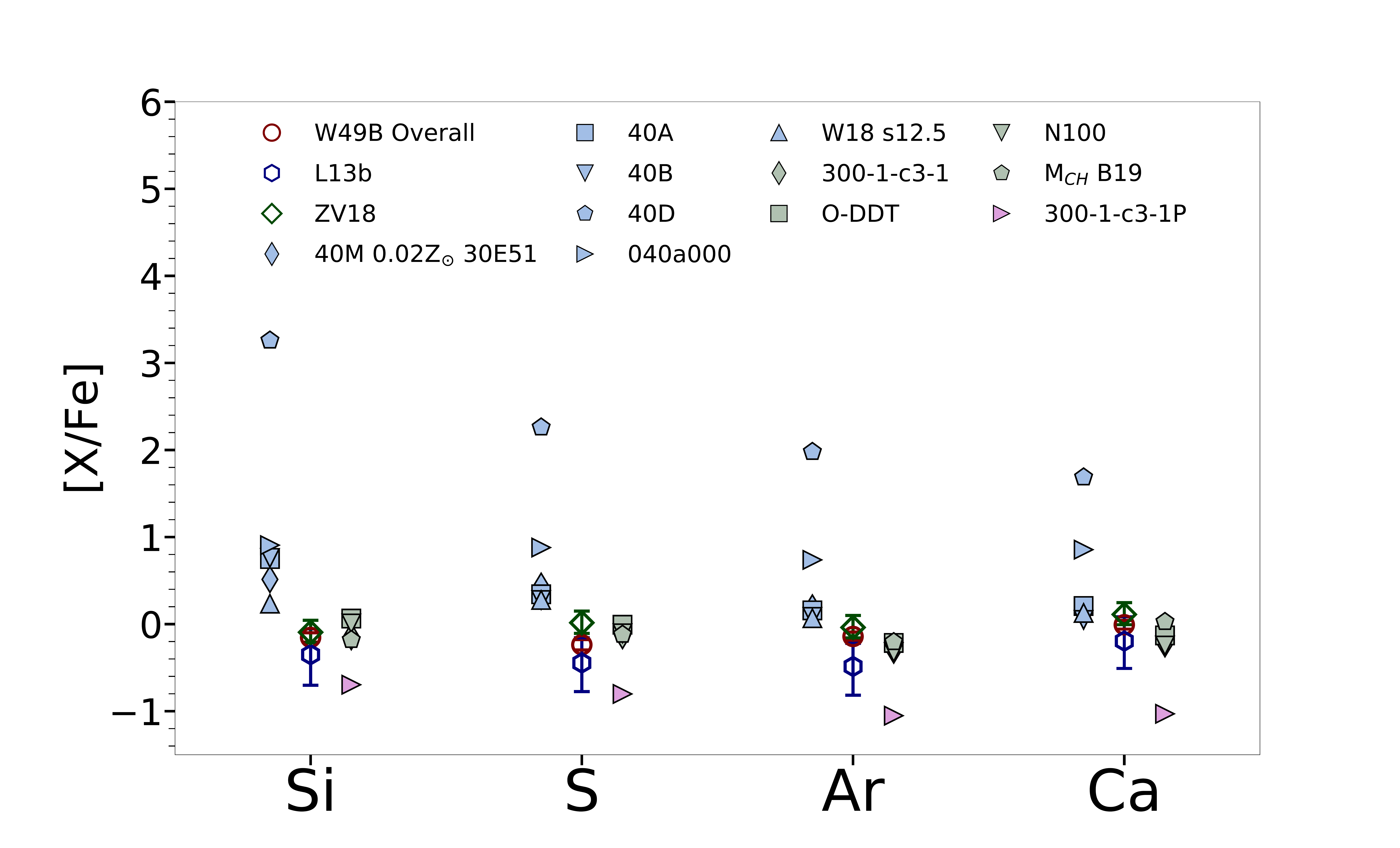}
\caption{The EM-weighted median logarithmic abundance ratio relative to Fe derived using SPI (\textit{red circle}), in comparison with the abundance values of  L13b (\textit{blue hexagon}) and ZV18 (\textit{green diamond}), as well as the selection of SN models from Table \ref{table:models}, plotted such that the core-collapse (\textit{light blue}) are on the left column and Type Ia (\textit{light green}) are on the right column. Pure deflagration Type Ia models are shown in \textit{pink}. For convenience, we shorten M$_{CH}$ $\xi=$0.9 $\rho_{DDT}$=2.8 $\times$ 10$^7$ $Z=$9.00 $\times$ 10$^{-3}$ to M$_{CH}$ B19.}
\label{fig:abund_comp}
\end{figure}

In Figure \ref{fig:abund_comp}, we present the EM-weighted median abundance values from SPI, compared to the selection of SN models, as well as the data from L13b and ZV18. The values are plotted relative to the Fe abundance, normalized to the appropriate solar value given by \citet{angr}.

The abundance values from L13b, ZV18, and SPI are found to be in agreement. The comparison of core-collapse models with the observations suggests that, relative to Fe, the core-collapse models have abundance values exceeding the observations. Furthermore, no single core-collapse model is compatible with the measured abundance of all elements.

The Type Ia models show much better agreement with the observations, with DDT models doing a better job than pure deflagration models. The near-Chandrasekhar mass model from \citet{bravo2019} provides the best fit, but other DDT models are not incompatible, given the uncertainties.

\begin{figure}[htbp]
\begin{center}
\includegraphics[width=\columnwidth]{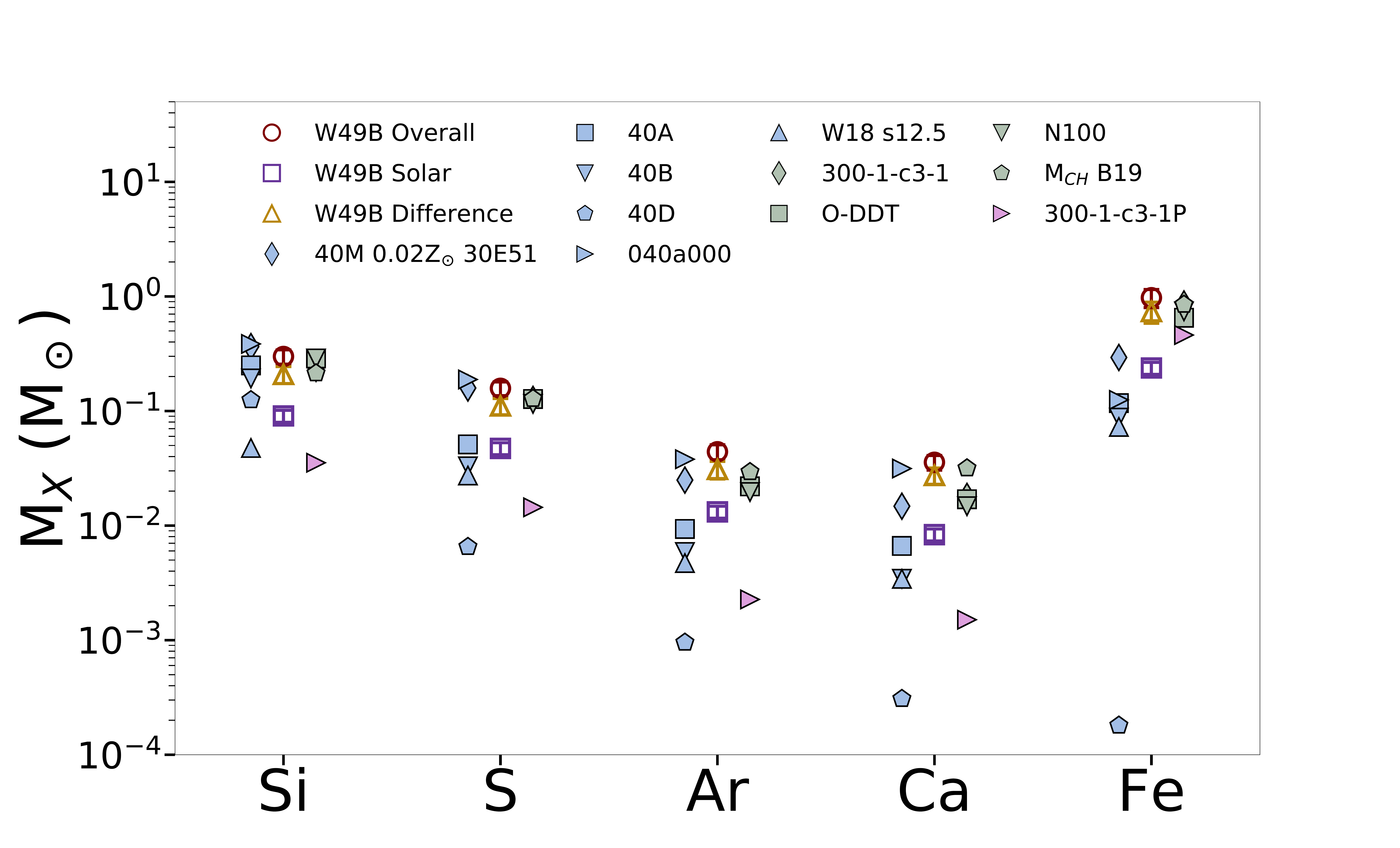}
\caption{The mass of each element over the entire remnant derived using SPI abundances (\textit{red circle}) and solar abundances (\textit{purple square}), as well as the difference between the two (\textit{gold triangle}), in comparison with the selection of SN models listed in Table \ref{table:models} (all in terms of solar mass). The models are represented using the same color and symbols as in Figure \ref{fig:abund_comp}. For convenience, we shorten M$_{CH}$ $\xi=$0.9 $\rho_{DDT}$=2.8 $\times$ 10$^7$ $Z=$9.00 $\times$ 10$^{-3}$ to M$_{CH}$ B19.}
\label{fig:yield_comp}
\end{center}
\end{figure}

We can also compare the mass of individual species, as calculated in \S \ref{sec:mass_den}, to that expected from a model. In Figure \ref{fig:yield_comp}, we plot for each element the mass inferred using SPI measured abundances and that found assuming solar abundances, as well as the difference between the two (as shown in Table \ref{table:yields}). This allows us to compare the observed yields with model predictions, while investigating the effect of possible contributions from the swept-up medium. As shown in Table \ref{table:yields}, we find that all masses are higher than that expected from a swept-up ambient medium with solar composition. These inferred masses are likely the products of the SN explosion, and it is reasonable to compare them with the predicted yields from the selected models. 

We find that the Type Ia DDT models provide the best match with the inferred masses. For Si, S and Ar there is considerable overlap between the core-collapse and Type Ia model predictions. These elements therefore have little discriminating power. Ca is also consistent with both core-collapse and Type Ia models, but with a greater degree of variation. The hypernova model of \citet{nomotoetal06} and the bipolar model of \citet{mn03} provide the best fit among the core-collapse models, but predict lower values than those observed. Fe has the greatest discriminating power, with all the Type Ia DDT models matching the inferred values, while core-collapse models are unable to do so. Similarly, the pure deflagration model fails in comparison to the DDT models. 

The conclusion that the inferred mass, from SPI abundances, is significantly larger than the inferred swept-up medium component, is based on the assumption that the surrounding medium has solar abundances (a common assumption for Galactic interstellar material). However, if the medium was formed by mass-loss from a core-collapse progenitor, the abundance values would resemble those in a stellar wind. In this case, the inferred mass for the elements in the swept-medium would increase, due to enrichment from the wind. This does not significantly affect the conclusions in Figure \ref{fig:yield_comp}, because a  wind would primarily show higher values of the lower Z elements, none of which are particularly discriminating. More importantly, wind material does not generally contain Fe, which is the main discriminant.

\begin{figure}
\gridline{
\includegraphics[width=\columnwidth, clip, trim=0.0cm 0.0cm 0.0cm 0.0cm]{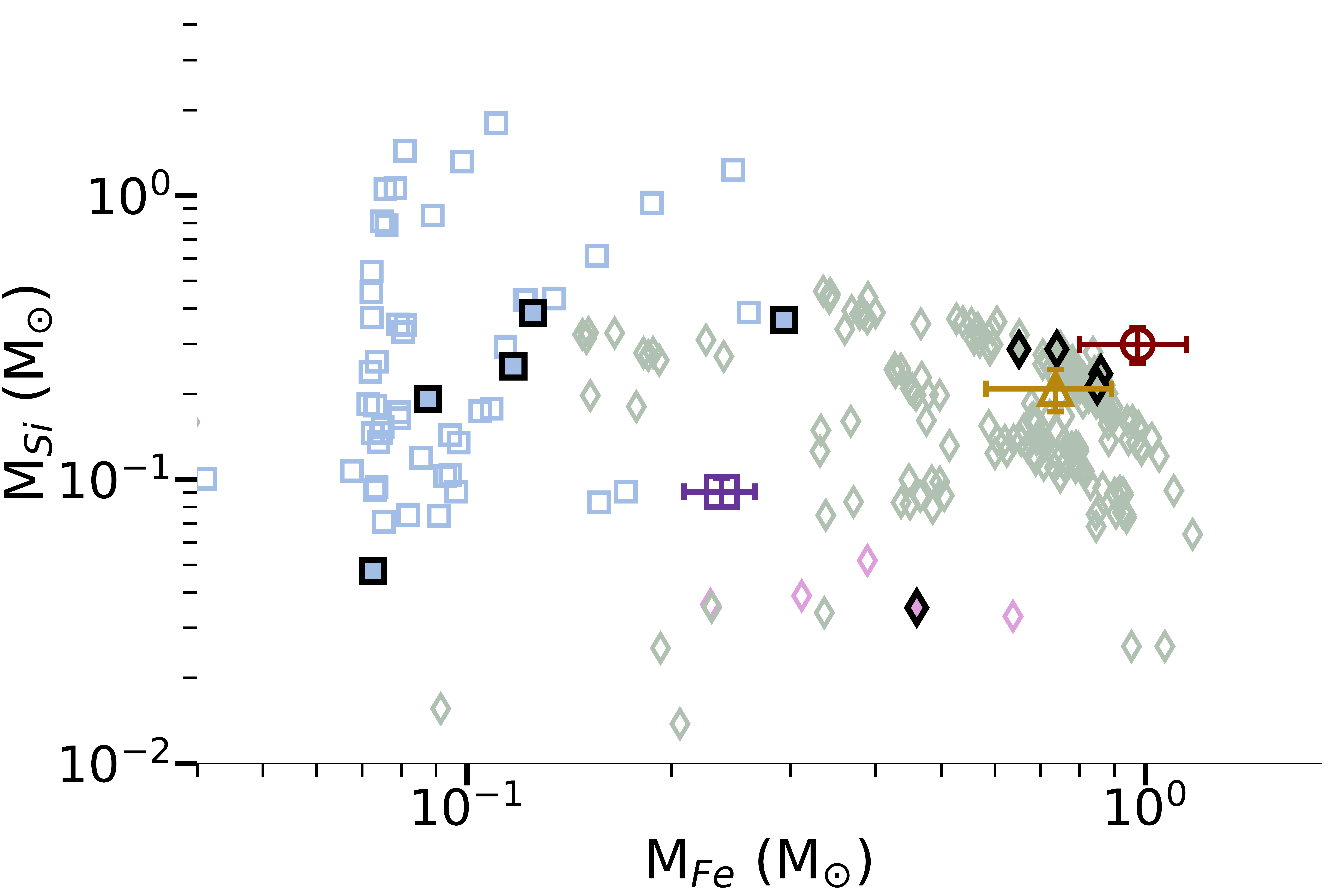}}
\gridline{
\includegraphics[width=\columnwidth, clip, trim=0.0cm 0.0cm 0.0cm 0.0cm]{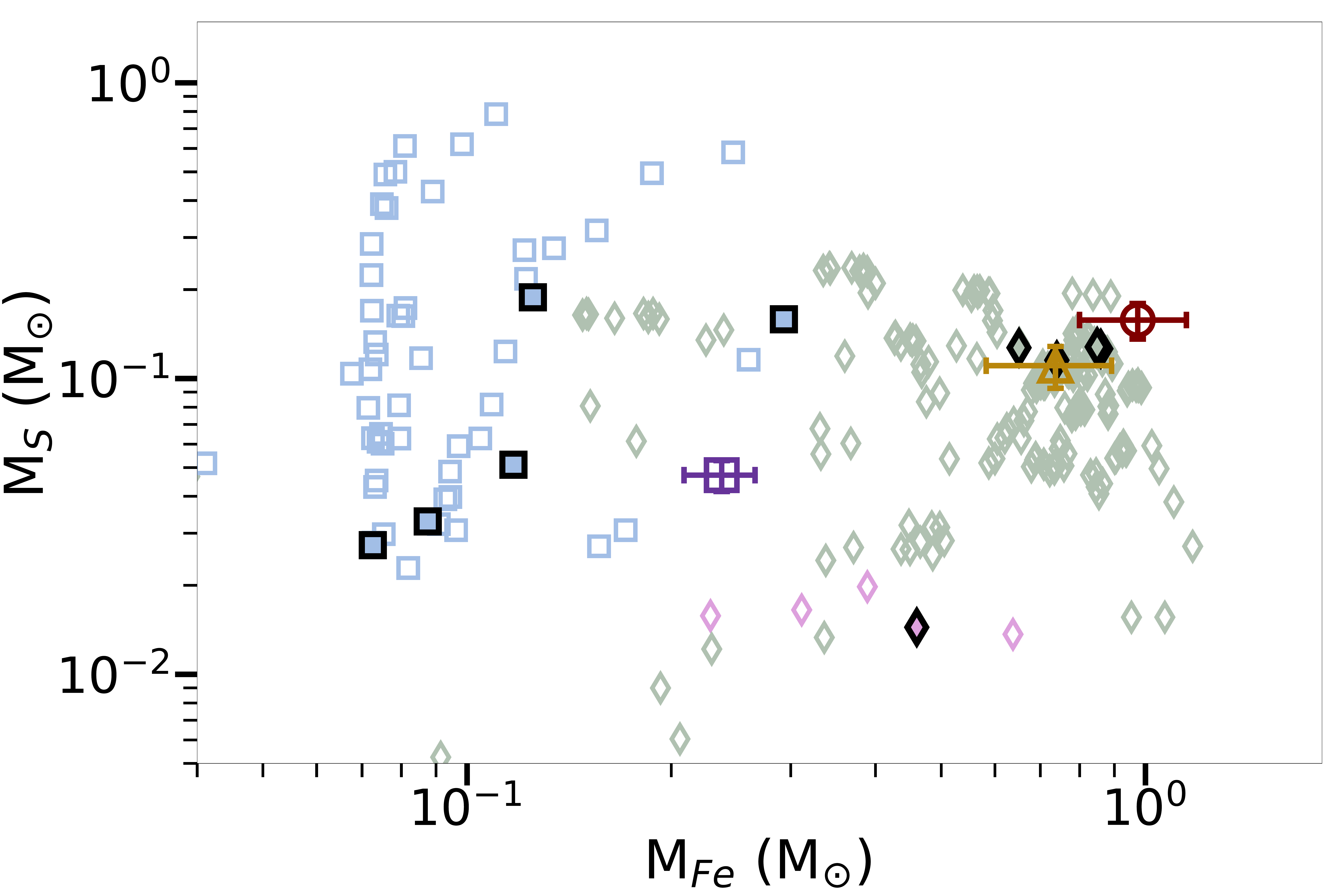}}

\caption{The mass of Si, S, and Fe, and over the entire remnant derived using SPI abundances (\textit{red circle}) and solar abundances (\textit{purple square}), as well as the difference between the two (\textit{gold triangle}) in comparison to the yield of all core-collapse models (\textit{light blue squares}) and all Type Ia models (\textit{light green and pink diamonds}) from the references in Table \ref{tab:families}. The filled and bold points represent the models selected for detailed comparison listed in Table \ref{table:models}.
}
\label{fig:multicomp}
\end{figure}

To investigate the robustness of the results in Figure \ref{fig:yield_comp}, we plot the Fe yield against Si and S for every SN model listed in the references of Table \ref{tab:families} (Figure \ref{fig:multicomp}). We again find that Type Ia models are the most compatible with our observations and that the DDT models provide a better match than the pure deflagration models.

\section{Assessing the Origin of W49B}

We review various characteristics of W49B that could provide information on the nature of its progenitor. Over the years, these have been used to attribute either a core-collapse or a Type Ia origin for the W49B progenitor. 

\subsection{Compact Central Object}

The core-collapse of a massive star is expected to leave a central object remnant, either a neutron star or black hole. Pulsations have not been detected \citep{gorham1996}, and no clear evidence of a compact central object has been found in W49B. L13b estimated an upper limit on the luminosity of an (undetected) neutron star, and compared that to known neutron star luminosities and model predictions of neutron star cooling. Their analysis suggested that if a compact object were to exist, it must be a black hole. ZV18 compared cooling neutron star models to point sources detected within the remnant. They find that potentially 9 of the sources are within the range of predicted luminosities for an age of 5-6 kyr. The lack of a detection leaves the debate about the origin unresolved, since it could mean that either  no central compact object exists; that one does exist but is still undetected; or that it could have collapsed directly to a black hole.

\subsection{Fe K$\alpha$ Centroid}

\citet{yamaguchi2014} measured the centroid of the Fe K$\alpha$ emission line in various remnants using \textit{Suzaku}. They found that Type Ia SNRs all have an Fe K$\alpha$ centroid at an energy $<$6550 eV, whereas core-collapse SNRs have centroids at higher energies. They attributed this distinction primarily to the ambient density into which these SNRs were expanding; Type Ia SNRs expand into a lower density medium, whereas core-collapse SNRs are initially expanding into much higher densities, created by mass-loss from the progenitor star. The location of the Fe, and how quickly it is shocked, affects the ionization state of the Fe, which in turn affects the position of the line centroid. They measure a centroid energy of $6663^{+1}_{-1}$ eV for W49B, which places it squarely in the core-collapse range. 

To investigate the position of the Fe K$\alpha$ centroid, we extracted the EPIC-pn 5.5-10 keV spectrum and fit it using an absorbed power-law model and Gaussian components for the line emission. We find the Fe K$\alpha$ centroid for the entire remnant is $6665^{+1}_{-1}$ eV, similar to the value reported by \citet{yamaguchi2014}. 

An alternative to the core-collapse explanation would be a Type Ia expanding in a dense medium, such as the interaction of W49B with a high density medium towards the east (\S \ref{sec:surr}). The interaction would lead to a shock being pushed into the dense medium, and a reflected shock traversing back into the SNR \citep{dwarkadas05}. This reflected shock would then ionize the Fe, leading to a high charged state and shifting of the centroid to higher energies.

A Type Ia SN expanding into a dense medium from an early time may also result in an Fe K$\alpha$ centroid at high energies.  While Type Ia SNe are expected to generally expand in unperturbed, low density ISM, there are some exceptions to this. The recent discovery of X-ray emission from a Type Ia-CSM SN \citep{bocheneketal18} suggests that there may be a class of rare and unusual Type Ia's that do harbor a very dense medium close in, although it is unclear how far this medium extends.  Other scenarios such as the core-degenerate scenario for Type Ia SNe \citep{soker13} also result in a high-density medium around the SN. Type Ia SNe expanding in such high density medium would shift the Fe K$\alpha$ centroid to higher energies.

\citet{patnaudeetal15} investigated the relationship between the Fe K$\alpha$ line luminosity and circumstellar density,  by modelling the expansion of core-collapse SNRs into a circumstellar medium formed by stellar mass-loss. They found that  W49B's Fe K$\alpha$ line luminosity was an anomaly, labelling it as an outlier for being much higher than those of core-collapse remnants. Expansion into a high density medium, no matter how or when in the SN evolution it happened, would also result in a larger Fe K$\alpha$ line luminosity, since thermal X-ray emission is proportional to the square of the density.

\subsection{Presence of a Wind-blown Bubble}
\label{sec:bubb}

Several studies have invoked a relationship between the progenitor mass and the characteristics of the progenitor's wind-blown bubble to assess the origin of W49B.  In their survey of SNRs observed with Spitzer, \citet{R06} noted that W49B showed [FeII] loops/filaments with H$_2$ emission on the edges, which they interpreted as indicating ionic shocks within a molecular environment predominantly on the E/SE side. \citet{K07} reported co-axial rings of [Fe II] emission, which they interpreted as enhanced density structures, such as those seen in wind-blown cavities around massive stars.  They concluded that the SNR had a massive star progenitor, and exploded as a core-collapse SN within a  wind-blown cavity that was bounded on one side by a molecular cloud. ZV18 related the size of the bubble to the progenitor mass, determining that for W49B the small size of the bubble implies either a CC origin with a progenitor mass of $\lesssim13$ \msun,  or a Type Ia origin.
    
Many of these claims are based on circumstantial evidence. While it is true that ring nebulae are seen around Wolf-Rayet stars, as cited by \citet{K07}, they are seen in far greater numbers around low-mass stars \citep{pnatlas}. The observation of rings by itself  does not point to a massive star; it could be a low mass star, or it may not be a star at all. As suggested by \citet{Z14}, the rings could also demarcate accumulated material in the binary plane. 

The existence of a wind-bubble, even if confirmed, does not uniquely suggest a massive progenitor. If Type Ia SNe arise from single degenerate progenitors, then it is possible that the wind from the companion may create a wind-bubble. Certainly, a wind-bubble has been proposed to explain the peculiar morphology of RCW86 \citep{williamsetal11}. A stellar wind bubble has also been proposed to explain the morphology of Kepler's SNR \citep{chiotellisetal12}. Therefore it is difficult to conclude much, if anything, about the progenitor type from the presence of a wind bubble, assuming one does indeed exist.

\subsection{Remnant Morphology}

Employing the power-ratio method outlined in \citet{lopezetal2009}, \citet{lopez2011} quantified the morphology of 24 SNRs observed by \textit{Chandra}, and found that W49B has a morphology consistent with a core-collapse SNR. However they also found W49B to be a singular exception, in that while variations in the power-profiles of different remnants were minor in all cases, W49B had significant variations between the power ratios of the Fe XXV and Si XIII lines. They attributed this to the case that W49B was the remnant of a jet-driven GRB explosion. We also note that \citet{lopez2011} limited their analysis to remnants not interacting with a molecular cloud; however, our results and others suggest that W49B is interacting with a dense medium (see \S \ref{sec:surr}). Given the distinction between the W49B results and those of all other SNRs, combined with the fact that W49B abundances are inconsistent with the few GRB models that are available, assumptions regarding the morphology of W49B are not particularly useful to constrain the progenitor. 

\section{Conclusions}
In this paper, we apply the Smoothed Particle Inference technique to supernova remnant W49B. Using the unique capabilities of SPI, we characterize the emitting plasma across the entire remnant and infer the density structure and mass of individual elements.

We find super-solar abundances for Si, S, Ar, Ca, and Fe. We also identify a strong temperature gradient from the southwest to the northeast, and find the Fe abundance to be elevated across the entire remnant. 

Using the inferred density in each blob, we compute the remnant's density structure and estimate the mass. In combination with multi-wavelength observations of W49B, the density structure suggests W49B is interacting with a dense medium towards the E/SE direction. We find a total mass of 130 $(\pm 16)$ {\msun} and an estimated ejecta mass of 1.2 $(\pm 0.2)$ {\msun}. Considering the uncertainties in our estimate, this ejecta mass is consistent with either a Chandrasekhar or sub-Chandrasekhar mass Type Ia progenitor. 

We compare the abundance ratios obtained with SPI to those of \citet{lopezetal13b}, \citet{zv18}, and a wide selection of core-collapse and Type Ia SN models. We find general agreement between the observed values among different authors, suggesting that past disagreements were due to model selection. DDT Type Ia models are found to provide the best match. Comparing instead the inferred mass yields from SPI, DDT Type Ia models are again found to be the most compatible.

While the abundance pattern suggests a Type Ia origin for W49B, we also review other remnant characteristics that could help identify the progenitor:
\begin{itemize}
    \item \textbf{Compact Central Object:} There is no clear evidence of a compact central object in W49B. Detection of a compact central object would be a definitive sign of a core-collapse origin, but non-detection remains inconclusive. 
    \item \textbf{Fe K$\alpha$ Centroid:} 
    The high centroid energy is taken as an indicator of a core-collapse origin. However the environment of W49B, especially the presence of a high density medium towards the E/SE, could be responsible for the centroid position. For a Type Ia scenario to be compatible with the high Fe K$\alpha$ centroid energy, it would be necessary
    to explain why it  would be evolving in a high-density region, which is rare, but not unknown for Type Ia SNe.
    \item \textbf{Presence of a Wind-blown Bubble:} The evidence for a wind-blown bubble is inconclusive. Furthermore, the presence of a wind-bubble does not necessarily denote a massive-star progenitor, as wind-blown bubbles may be formed around Type Ia SNe by a companion star, such as proposed for RCW86.
    \item \textbf{Remnant Morphology:} \citet{lopezetal13b} associated the morphology of W49B with a GRB progenitor, but they found several inconsistencies. The abundance values are not consistent with a GRB, and the morphology may have been affected by the presence of a nearby dense medium.
\end{itemize}

In summary, given the complicated and ill-understood relationship between the progenitor's mass loss, environment, morphology, and the SNe explosion, as demonstrated by the ever-growing collection of unusual SNe, it is clear that these factors cannot be reliably used to infer the nature of W49B's progenitor. While the abundance of Fe suggests a Type Ia origin, other properties remain inconclusive. We stress that a complete model of the remnant must be able to explain the density structure of W49B (Figure \ref{fig:general_maps}), the spatial abundance distribution, including the presence of strong Fe emission throughout (Figure \ref{fig:mass_distr}), and the relative abundances (Figure \ref{fig:yield_comp}). Further study, perhaps involving hydrodynamical simulations to model the environment of W49B, and extended comparison to forthcoming SN explosion models, is clearly warranted.    

\ \\
{\bf Acknowledgments:}
We thank the referee for their comments, which have helped to substantially improve the paper. This work was partially supported by NASA ADAP grant NNX15AF03G to Pennsylvania State University, with subcontracts to the University of Chicago and Northwestern University. VVD is also supported by NSF grant 1911061. Based on observations obtained with XMM-Newton, an ESA science mission with instruments and contributions directly funded by ESA Member States and NASA.

\facilities{XMM(EPIC)}

\bibliography{spipaper}%

\end{document}